\documentclass[fp]{jpsj3}
\usepackage{txfonts}
\usepackage{multirow}
\usepackage{bm}
\usepackage{amsmath}
\usepackage{pdfpages}
\usepackage{mathrsfs}
\usepackage{braket}
\usepackage{physics}

\makeatletter
\newcommand\newblock{\hskip .11em\@plus.33em\@minus.07em}
\makeatother

\title{Photon-Assisted Tunneling in Double Quantum Dot: Application of Scattering Theory}

\author{Miyu Umebayashi and Mikio Eto\thanks{eto@rk.phys.keio.ac.jp}}
\inst{Faculty of Science and Technology, Keio University, Hiyoshi,
Yokohama 223-8522, Japan} 

\abst{
We theoretically examine the photon-assisted tunneling (PAT) in a double quantum dot
(DQD) in parallel when one of the quantum dots (QDs) is irradiated by an AC field.
First, we formulate the PAT in a single QD by solving the time-dependent
Schr\"odinger equation using the scattering theory. The QD has an
oscillating energy level,
$\varepsilon(t)=\varepsilon_0+eV_{\mathrm{AC}}\cos\omega t$,
and is connected to two leads by the tunnel coupling $\Gamma$.
We show that the resonant tunneling takes place
through energy levels of the polariton, $\varepsilon_0+N\hbar\omega$
($N=0,\pm 1, \pm 2, \cdots$), when $\Gamma \ll \hbar\omega$ (PAT) and
through the energy level $\varepsilon(t)$
when $\Gamma \gg \hbar\omega$ (adiabatic transport). 
Then, the scattering theory is applied to the PAT in the DQD
in the presence of magnetic flux penetrating between the QDs.
We observe the Aharonov--Bohm effect not only in the main peak ($N=0$) but
also in subpeaks ($N \ne 0$), indicating coherent transport through the
polariton states. Our theory is also applicable to the DQD in the three-terminal geometry.
We demonstrate the phase measurement through the irradiated QD and show that the
measured phase shift changes continuously from 0 to $\pi$ around both
the main peak and subpeaks.}

\begin{document}
\maketitle

\section{Introduction}
The single electron tunneling accompanied by photon emission or absorption
is called photon-assisted tunneling (PAT). It was first reported by Dayem and
Martin in 1962 in an experiment on a superconducting film between two insulators
in a microwave field.\cite{dayem_quantum_1962}
In the following year, Tien and Gordon described the PAT in the superconducting diode
using a model of an oscillating energy
level.\cite{Tien-Gordon, platero_photon-assisted_2004}

In the study of mesoscopic physics,
the transport through a semiconductor quantum dot (QD) in an AC field is one of the
important issues  and has been studied for a long time.
There are two extreme cases concerning
the strength $\Gamma$ of tunnel coupling to external leads and the frequency $\omega$
of the AC field.\cite{Kouwenhoven}
The PAT was observed in the case of $\Gamma \ll \hbar \omega$:
When the energy level is given by
\begin{equation}
F_{\mathrm{AC}}(t)=\varepsilon_{0}+eV_{\mathrm{AC}}\cos \omega t
\label{eq:F_AC}
\end{equation}
in the QD, the conductance shows a main peak when the Fermi level
$E_{\mathrm F}$ matches
$\varepsilon_0$ and subpeaks when $E_{\mathrm F}=\varepsilon_0+N\hbar\omega$
($N=\pm 1, \pm 2, \cdots$).\cite{kouwenhoven_photon-assisted_1994_257,
kouwenhoven_observation_1994_256, blick_photonassisted_1995,
fujisawa_photon_1997, oosterkamp_photon_1997, Oosterkamp1998,
kawano_terahertz_2008, shibata_photon-assisted_2012, yoshida_terahertz_2015}
This indicates the transport through the QD with the emission
or absorption of $|N|$ photons. In other words, the transport takes place
through the polariton states of the energy levels
$\varepsilon_0+N\hbar\omega$ in the QD.
The experimental results are well explained by the Tien--Gordon theory.
In the opposite case of $\Gamma \gg \hbar \omega$, adiabatic transport takes
place through the energy level $\varepsilon(t)$. A notable example
of the adiabatic transport is the single-electron turnstile applied to
the current standard.\cite{Geerligs1990,Kouwenhoven1991,
Pekola2013,Kaestner2015}

Lately, the PAT was observed by irradiating terahertz (THz) light
on an InAs QD\cite{shibata_photon-assisted_2012} or a single molecule of
$\mathrm{C}_{60}$.\cite{yoshida_terahertz_2015}
The light is focused on the QD beyond the diffraction limit.
A sensitive detector of the THz light was proposed, utilizing the interlevel transition
in an InAs QD.\cite{zhang_terahertz_2015}
More generally, the interaction between an electron in the QD and
a photon, e.g., cavity QED, is being studied extensively for quantum information
processing.\cite{Walther2006,Reiserer2015}
Therefore, deep understanding of the PAT and other transport
phenomena in an AC field will be required for the application of
the QD to future technology.

In this paper, we theoretically examine the transport through a single QD
and a double quantum dot (DQD) in parallel when the energy level is
given by Eq.\ (\ref{eq:F_AC}).
We disregard the electron--electron interaction $U$ in the QDs.
We do not expect $U$ to cause a qualitative change, such as the Kondo effect,
in the vicinity of the electric current peaks mainly studied in this paper.
We solve the time-dependent Schr\"odinger equation
using the scattering theory\cite{Hewson1993,JJSakurai2021}
for the following reasons:

(i) First, we reformulate the electric current through a single QD in both
the cases of PAT and adiabatic transport. Although the current expression
in the former was previously derived using the Keldysh nonequilibrium
Green's function method\cite{jauho_time-dependent_1994} or
others,\cite{platero_photon-assisted_2004,kohler_driven_2005}
our tutorial method intuitively
illustrates the physical pictures of the time-dependent transport phenomena
for both $\Gamma \ll \hbar \omega$ and $\Gamma \gg \hbar \omega$.

(ii) In the DQD in parallel, we assume that one of the QDs has the oscillating
energy level in Eq.\ (\ref{eq:F_AC}) and the other has a constant
energy level. The infinite series of the perturbation with respect to the tunnel
coupling to the leads enables the calculation of the electric current
to the second-order in $eV_{\mathrm{AC}}/(2\hbar \omega)$.
In the presence of magnetic flux penetrating between the QDs,
we elucidate the Aharonov--Bohm (AB) interference effect on
the transport through the polariton states
when the AC field is considered using Eq.\ (\ref{eq:F_AC}) in a classical way.
Note that the AB oscillation in the DQD in the absence of the AC field
has been studied experimentally\cite{holleitner_coherent_2001, hatano_gate-voltage_2004}
and theoretically.\cite{kubala_flux-dependent_2002, bai_effect_2004, lu_fano_2006,
tokura_interference_2007, li_coulomb_2009, hartle_decoherence_2013}.
In the case of its presence, however, few works have been reported.

(iii) The scattering theory is also applicable to the DQD in
the three-terminal geometry. We demonstrate the phase measurement through
the QD with the oscillating energy level
using the other QD as a reference arm of the
``double slit interference experiment''.\cite{zhang_kondo_2022}
The phase measurement is impossible in the transport experiment
in the two-terminal geometry owing to the restriction of
Onsager's reciprocal theory.\cite{yacoby_coherence_1995}
The measurement of the phase shift through a QD was reported using
a small ring with an embedded QD in a three- or four-terminal geometry
in the absence of the AC field.\cite{schuster_phase_1997,
van_der_wiel_kondo_2000, ji_phase_2000, Takada2014}
In its presence,
we show that the measured phase shift changes continuously from 0 to $\pi$ around both
the main peak and subpeaks in accordance with the Breit--Wigner resonance.
We do not observe the ``phase lapse'' between the main peak and
a subpeak, as reported previously between successive current
peaks without the AC field.\cite{schuster_phase_1997,
ji_phase_2000,Avinun-Kalish2005,Karrasch2007}

This paper is organized as follows.
In Sect.\ \ref{section:Model and Calculation Method},
we explain our models and formulation based on the time-dependent
Schr\"odinger equation. Section \ref{section:Calculated results1} is
devoted to the PAT and adiabatic transport through a single QD
using the scattering theory. We mention the relationship between our
Green's function and that of the Floquet theory.\cite{floquet_sur_1883,
eliasson_floquet_1992,ishizuka_local_2017}
In Sect.\ \ref{section:Calculated results2},
we derive the electric current through the DQD in parallel
when one of the QDs has an
oscillating energy level. In the two-terminal setup, we show the
AB effect on the transport through the polariton states.
In the three-terminal setup, we discuss the phase measurement
through the oscillating energy level.
The conclusions are given in Sect.\ \ref{section:Conclusion}. 

 \begin{figure}
  \centering
  \includegraphics*[width=0.45\textwidth]{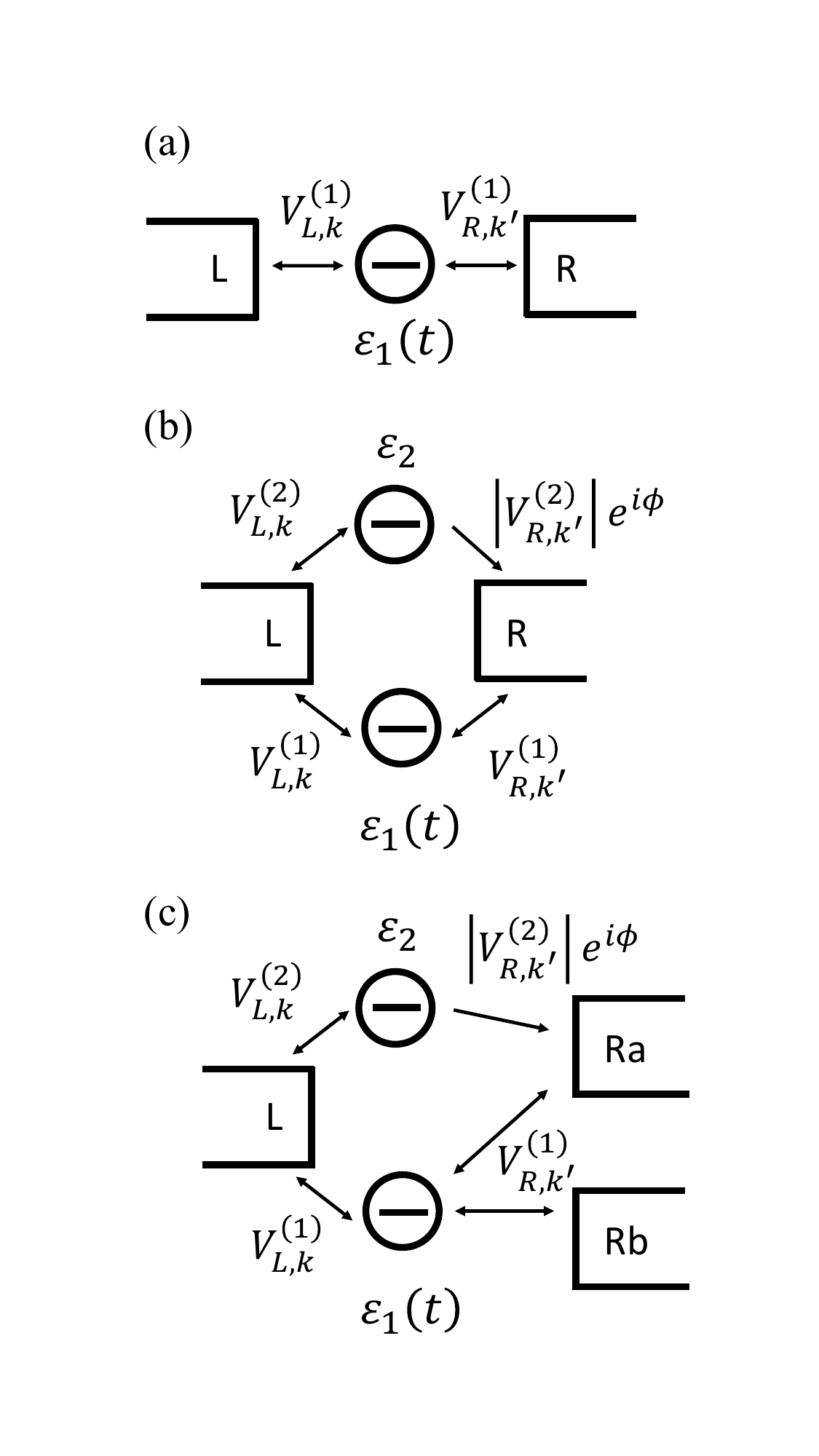}
  \caption{
  (a) Model for a single QD with an oscillating energy level
  $\varepsilon_1(t)=F_{\mathrm{AC}}(t)$ in
  Eq.\ (\ref{eq:F_AC}). The QD is connected to leads $L$ and $R$
  by the tunnel couplings $V_{L k}^{(1)}$ and $V_{R k'}^{(1)}$,
  respectively.
  (b, c) Model for a DQD in parallel in a two- or three-terminal
  geometry. The energy levels in the QDs are 
  $\varepsilon_1(t)=F_{\mathrm{AC}}(t)$ in QD1
  and time-independent $\varepsilon_{2}$ in QD2.
  QD$j$ is connected to leads $L$ and $R$
  by the tunnel couplings $V_{L k}^{(j)}$ and $V_{R k'}^{(j)}$,
  respectively. $V_{R k'}^{(2)} = \left| V_{R k'}^{(2)} \right| e^{i\phi}$,
  where $\phi=2\pi\Phi/(h/e)$ is the AB phase for the magnetic
  flux $\Phi$ penetrating between the QDs. 
  In panel (c), lead $Rb$ is connected to QD1 [by $V_{R k'}^{(1)}$
  when state $k'$ belongs to the lead] but not to QD2.
  }
  \label{fig:model}
\end{figure}

\section{Model and Calculation Method}
\label{section:Model and Calculation Method}

\subsection{Model}
\label{subsection:Model}

We examine three models depicted in Fig.\ \ref{fig:model}; single QD with an
energy level $\varepsilon_1(t)=F_{\mathrm{AC}}(t)$ in Eq.\ (\ref{eq:F_AC}),
DQD with the oscillating energy level in one of the QDs and
a static level in the other in two- and three-terminal geometries.
We describe the Hamiltonian for the DQD model in Fig.\ \ref{fig:model}(b)
in this subsection. For a single QD in Fig.\ \ref{fig:model}(a),
we set $\varepsilon_2=0$ in Eq.\ (\ref{eq:Hdot})
and $V_{L k}^{(2)}=V_{R k'}^{(2)}=0$ in Eq.\ (\ref{eq:HT-DoubleQD}).
For the three-terminal situation in
Fig.\ \ref{fig:model}(c), we divide the summation over $k$ in lead $R$ into
those in leads $Ra$ and $Rb$ in Eqs.\ (\ref{eq:Hlead}) and
(\ref{eq:HT-DoubleQD}).

The Hamiltonian is given by
 \begin{equation}
  H(t) = H_{\mathrm{dot}}(t) + H_{\mathrm{leads}} + H_{\mathrm{T}},
  \label{eq:H}
 \end{equation}
where
 \begin{align}
  H_{\mathrm{dot}}(t)
  &= \varepsilon_1(t) d_1^{\dag}d_1+\varepsilon_2 d_2^{\dag}d_2,
  \label{eq:Hdot}\\
  H_{\mathrm{leads}}
  &= \sum_{\alpha=L,R} \sum_{k} \varepsilon_{k} a^{\dag}_{\alpha k} a_{\alpha k},
  \label{eq:Hlead}\\
  H_{\mathrm{T}}
  &= \sum_{\alpha=L,R} \sum_{k} \sum_{j=1,2}
  \left(V_{\alpha k}^{(j)} a^{\dag}_{\alpha k} d_{j} + \mathrm{h.c.} \right).
  \label{eq:HT-DoubleQD}
 \end{align}
Here, $d^{\dag}_j$ and $d_j$ are the creation and annihilation operators of
an electron in QD$j$ while $a^{\dag}_{\alpha k}$ and $a_{\alpha k}$ are those in
lead $\alpha$ ($=L$, $R$) with state $k$.
The energy levels in the QDs are given by $\varepsilon_1(t)=F_{\mathrm{AC}}(t)$
and time-independent $\varepsilon_2$.
The spin of the electrons is omitted in this paper.

In the tunnel Hamiltonian $H_{\mathrm{T}}$,
QD$j$ is connected to state $k$ in lead $\alpha$ by $V_{\alpha,k}^{(j)}$.
The magnetic flux $\Phi$ penetrating between the QDs is taken into account by
the AB phase $\phi=2\pi\Phi/ \left( h/e \right)$.
We choose the gauge in such a way as
$V_{R k'}^{(2)} = \left| V_{R k'}^{(2)} \right| e^{i\phi}$ and
$V_{L k}^{(1)}, V_{L k}^{(2)}$, and $V_{R k'}^{(1)}$ are real and positive. 

In the case of a single QD, the strength of the tunnel coupling to lead $\alpha$
is characterized by the linewidth,
 \begin{equation}
  \Gamma^{\alpha} (\varepsilon) = 2\pi\sum_{k} \left| V_{\alpha,k}^{(1)} \right|^2
  \delta \left( \varepsilon_{k}-\varepsilon \right).
  \label{eq:linewidth_1QD}
 \end{equation}
Assuming  that its dependence on $\varepsilon$ is weak
around the Fermi level $E_{\mathrm F}$, we simply denote
$\Gamma^{\alpha}$ hereafter. The total linewidth in the QD is given by
 \begin{equation}
  \Gamma=\Gamma^{L}+\Gamma^{R}.
  \label{eq:totallinewidth_1QD}
 \end{equation}

In the DQD in the two-terminal setup, we define the linewidth function
$\mathbf{\Gamma}^{\alpha}$ in a matrix form of $2\times2$ corresponding to QD1
and QD2 by
 \begin{equation}
  \Gamma_{ij}^{\alpha} = 2\pi\sum_{k}
  \left[ V_{\alpha,k}^{(i)} \right] ^{\ast}V_{\alpha,k}^{(j)}
\delta \left( \varepsilon_{k}-\varepsilon \right)
  \label{eq:elementofGamma}
 \end{equation}
around $\varepsilon \simeq E_{\mathrm F}$.
The diagonal element $\Gamma_{jj}^{\alpha}$ indicates the
linewidth in QD$j$ due to the tunnel coupling to lead $\alpha$.
For the off-diagonal element in Eq.\ (\ref{eq:elementofGamma}) ,
we define $p_{\alpha}$ as
 \begin{equation}
  p_{\alpha} = \left| \Gamma_{12}^{\alpha} \right| /
   \sqrt{\Gamma_{11}^{\alpha} \Gamma_{22}^{\alpha}}
  \label{eq:p}
 \end{equation}
($0 \leq p_{\alpha} \leq 1$),
which represents the coherence between the
QDs through lead $\alpha$.\cite{kubo_p,zhang_kondo_2022}
$p_{\alpha}$ is identical to the overlap integral between the conduction
modes coupled to QD1 and QD2 in lead $\alpha$ at the energy
$\varepsilon \simeq E_{\mathrm F}$.\cite{Eto2020}
In the case of a single conduction channel in the lead,
$p_{\alpha}=1$ and the connection between the QDs is maximal.
With an increase in channel number, $p_{\alpha}$ usually decreases.
When $p_{\alpha}=0$, the two QDs are independent of each other.
In experiments, $p_{\alpha}$ is determined by the device structure.

The total linewidth function in the DQD is given by
 \begin{equation}
  \mathbf{\Gamma}=\mathbf{\Gamma}^{L}+\mathbf{\Gamma}^{R}.
  \label{eq:QD1Gamma1}
 \end{equation}

Similarly, we introduce the linewidth functions $\mathbf{\Gamma}^{L}$,
$\mathbf{\Gamma}^{Ra}$, and $\mathbf{\Gamma}^{Rb}$ for the model in
Fig.\ \ref{fig:model}(c). Since QD2 is disconnected from lead $Rb$,
$\mathbf{\Gamma}^{Rb}=\mathrm{diag}(\Gamma^{Rb}_{11},0)$ and $p_{Rb}=0$.

\subsection{Formulation}
\label{subsection:Calculation Method}

Using the time-dependent Schr\"{o}dinger equation,
 \begin{equation}
   i\hbar\frac{\partial}{\partial t} \Ket{{\psi} (t)} =
   [H_{\mathrm{dot}}(t) + H_{\mathrm{leads}}
   + H_{\mathrm{T}}] \Ket{{\psi} (t)},
  \label{eq:schrodingereq}
 \end{equation}
we perform the perturbative calculation with respect to $H_{\mathrm{T}}$.

For the DQD model in Fig.\ \ref{fig:model}(b),
the wavefunction is written as
\begin{equation}
  \Ket{\psi (t)} =  \sum_{j=1,2} C_j(t) e^{-i E_j(t)/\hbar} \Ket{d_j}
   + \sum_{\alpha=L,R} \sum_{k} C_{\alpha,k}(t)
      e^{-i \varepsilon_{k}(t-t_0)/\hbar} \Ket{\alpha,k},
  \label{eq:wavefn-DQD}
\end{equation}
where
$\Ket{d_j}=d_j^{\dag}\Ket{0}$ and
$\Ket{\alpha,k}=a^{\dag}_{\alpha k}\Ket{0}$
with $\Ket{0}$ being the vacuum state,
\begin{align}
E_1(t) & = \int_{t_0}^t \varepsilon_{1}(t') dt'
\label{eq:E_1org}
\\
& = \varepsilon_{0} (t-t_0) +
\frac{eV_{\mathrm AC}}{\omega}
(\sin \omega t - \sin \omega t_0),
\label{eq:E_1}
\end{align}
and $E_2(t)=\varepsilon_{2} (t-t_0)$.

The substitution of Eq.\ (\ref{eq:wavefn-DQD}) into
Eq.\ (\ref{eq:schrodingereq}) yields
Eqs.\ (\ref{eq:wavefn-DQD1}) and (\ref{eq:wavefn-DQD2})
in Appendix \ref{section:Calculation_suppl}.
From these equations, we derive the
transition rate to the lowest order in $H_{\mathrm{T}}$
 (Fermi's golden rule; Sect.\ \ref{section:Calculated results1a})
and the scattering matrix to the infinite order in $H_{\mathrm{T}}$
(Sects.\ \ref{section:Calculated results1b} and
\ref{section:Calculated results2}).

In the following sections, we denote the Fermi level in leads $L$
and $R$ ($Ra$, $Rb$) by $\mu_L$ and $\mu_R$ ($\mu_{Ra}$,
$\mu_{Rb}$), respectively.
$\mu_L-\mu_R=eV_{\mathrm{bias}}$ under the bias voltage
$V_{\mathrm{bias}}$.
We set $\mu_{Ra}=\mu_{Rb} \equiv \mu_R$ in Fig.\ \ref{fig:model}(c).
We change the original energy level $\varepsilon_0$ in the QD,
the first term on the right side of Eq.\ (\ref{eq:F_AC}),
which can be tuned by the gate voltage in experiments.

\begin{figure}
  \centering
  \includegraphics*[width=0.43\textwidth]{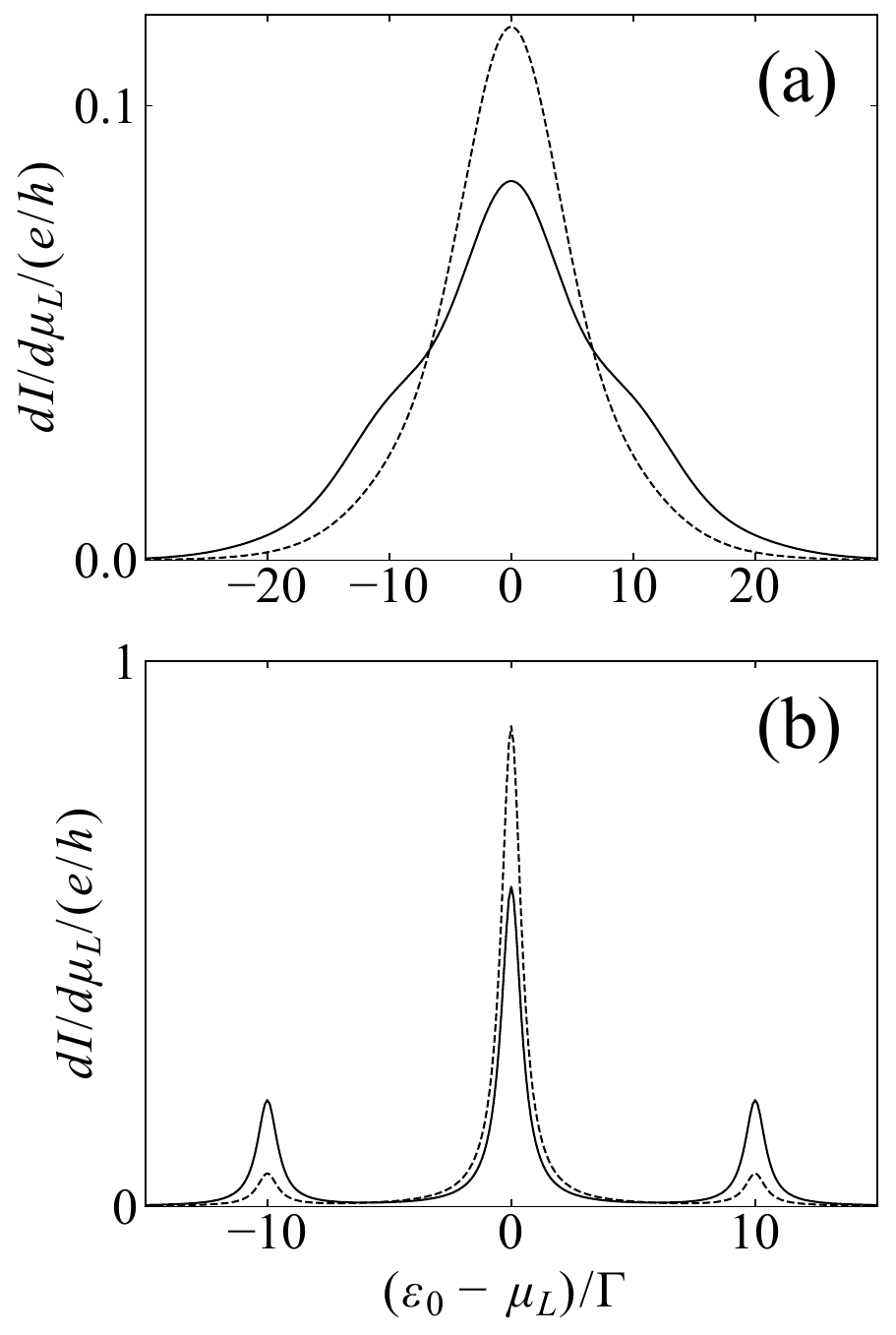}
  \caption{
  Differential conductance $d I_{L \rightarrow R}/d \mu_L$
  as a function of $\varepsilon_{0}$ for a single QD
  depicted in Fig.\ \ref{fig:model}(a), when
  $\Gamma \ll \hbar\omega$
  with $\Gamma$ being the linewidth due to the tunnel coupling to
  the leads ($\Gamma_L=\Gamma_R=\Gamma/2$).
  The energy level in the QD is
  $\varepsilon_1(t)=F_{\mathrm{AC}}(t)$ in Eq.\ (\ref{eq:F_AC}).
  The temperature is (a) $k_{\mathrm B} T \gg \Gamma$
  [$k_{\mathrm B} T/\Gamma=3$; Eq.\ (\ref{eq:seq_cond_1QD})]
  and (b) $k_{\mathrm B} T \ll \Gamma$
  [Eq.\ (\ref{eq:conductance_1QD2})].
  The frequency of the AC field is $\hbar\omega/\Gamma = 10$ and
  its amplitude is $eV_{\mathrm{AC}}/\Gamma = 5$ (broken line) and
  $10$ (solid line).
  }
  \label{fig:output_Single QD}
\end{figure}

\section{Single Quantum Dot}
\label{section:Calculated results1}

In this section, we examine the single QD model depicted in
Fig.\ \ref{fig:model}(a), by setting $V_{L k}^{(2)}=
V_{R k}^{(2)}=0$ and
$C_{2}(t)=0$ in Eqs.\ (\ref{eq:wavefn-DQD1}) and (\ref{eq:wavefn-DQD2}).
We begin with the sequential tunneling to the lowest order in
$H_{\mathrm{T}}$. Next, we obtain the exact formula of the electric
current through the summation of the perturbative series to the infinite order.

\subsection{Sequential tunneling}
\label{section:Calculated results1a}

First, we assume that the transport through the QD takes place sequentially
from a lead to the QD and from the QD to another lead. This assumption
will be verified at high temperatures of
$k_{\mathrm B}T \gg \Gamma$ in the next subsection.

We calculate the transition rate between the QD and leads to the
lowest order in $H_{\mathrm{T}}$ in
Appendix \ref{section:Calculation_suppl_1}. We examine the
two extreme cases of $\Gamma \ll \hbar \omega$ and
$\Gamma \gg \hbar \omega$.

\subsubsection{Case of $\Gamma \ll \hbar \omega$}

In the case of $\Gamma \ll \hbar \omega$, the transition rate
from the QD to lead $\alpha$ is given by
Eq.\ (\ref{eq:transition_rate_1QD}). This results in the
electric current
\begin{equation}
I_{L \rightarrow R}=
\frac{e}{\hbar}
\frac{\Gamma_L \Gamma_R}{\Gamma_L+\Gamma_R}
\sum_{N,N'} \left[ J_N(\tilde{V}_{\mathrm AC}) \right]^2
\left[ J_{N'}(\tilde{V}_{\mathrm AC}) \right]^2
[f_L(\varepsilon_0+N\hbar\omega)-f_R(\varepsilon_0+N'\hbar\omega)],
\label{eq:seq_current_1QD}
\end{equation}
where $\tilde{V}_{\mathrm AC}=eV_{\mathrm AC}/(\hbar\omega)$,
$J_N$ is the $N$th Bessel function, and
$f_{\alpha}(E)=[1+e^{(E-\mu_{\alpha})/(k_{\mathrm B}T)}]^{-1}$ is
the Fermi distribution function in lead $\alpha$ at temperature $T$.
This expression was obtained
previously.\cite{platero_photon-assisted_2004,kohler_driven_2005}.
The differential conductance
\begin{equation}
\frac{d I_{L \rightarrow R}}{d \mu_L} =
\frac{e}{\hbar}
\frac{\Gamma_L \Gamma_R}{\Gamma_L+\Gamma_R}
\frac{1}{4k_{\mathrm B} T}
\sum_N \left[ J_N(\tilde{V}_{\mathrm AC}) \right]^2
\cosh^{-2}
\frac{\varepsilon_0+N\hbar\omega-\mu_{L}}
{2k_{\mathrm B} T}
\label{eq:seq_cond_1QD}
\end{equation}
shows a peak structure:
A main peak at $\mu_{L}=\varepsilon_0$ and subpeaks
at $\mu_{L}=\varepsilon_0 + N\hbar\omega$ ($N = \pm 1,
\pm 2, \cdots$) with the weight of
$\left[ J_N(\tilde{V}_{\mathrm AC}) \right]^2$.
This indicates that the polariton states are formed
in the QD at energy levels $\varepsilon_0 + N\hbar\omega$.
Note that we do not include the factor $2$ for the spin
degrees of freedom in calculating the electric current and
differential conductance.

Figure \ref{fig:output_Single QD}(a) shows
$d I_{L \rightarrow R}/d \mu_L$ in Eq.\ (\ref{eq:seq_cond_1QD})
as a function of $\varepsilon_0$. Besides the main peak,
the subpeaks of $N=\pm 1$ appear in the shoulders of the main
peak. The peak widths are determined by the temperature
$k_{\mathrm{B}}T$.

\begin{figure}
  \centering
  \includegraphics*[width=0.43\textwidth]{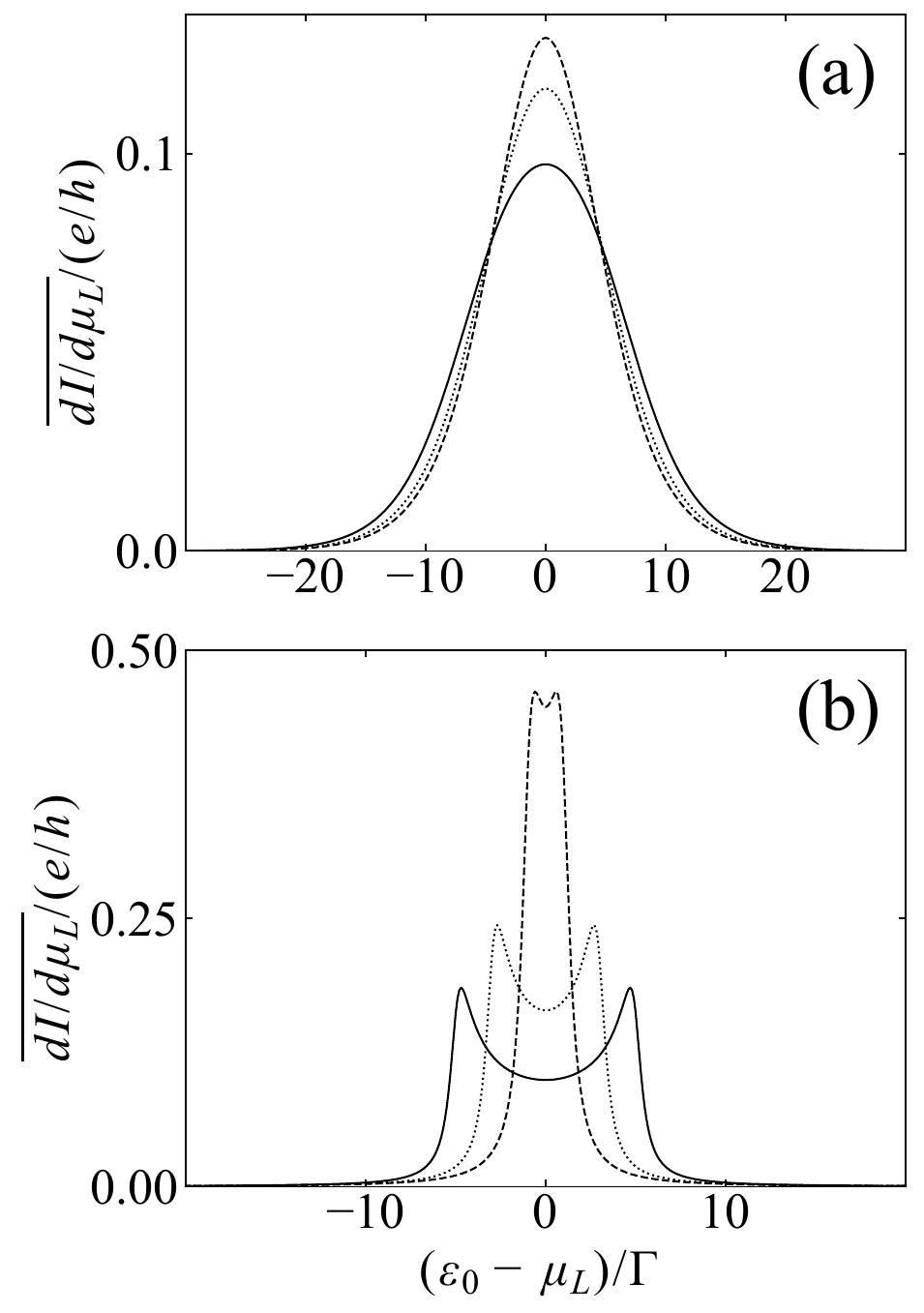}
  \caption{
  Time-averaged differential conductance
  $\overline{d I_{L \rightarrow R}/d \mu_L(t)}$
  as a function of $\varepsilon_{0}$ for a single QD
  depicted in Fig.\ \ref{fig:model}(a), when
  $\Gamma \gg \hbar\omega$
  with $\Gamma$ being the strength of  tunnel coupling to the leads
  ($\Gamma_L=\Gamma_R=\Gamma/2$).
  The energy level in the QD is $\varepsilon_1(t)=F_{\mathrm{AC}}(t)$
  in Eq.\ (\ref{eq:F_AC}).
  The temperature is (a) $k_{\mathrm B} T \gg \Gamma$
  [$k_{\mathrm B} T/\Gamma=3$; time-average of
  Eq.\ (\ref{eq:seq_cond_1QDb})]
  and (b) $k_{\mathrm B} T \ll \Gamma$
  [that of Eq.\ (\ref{eq:conductance_1QDb2})].
  The amplitude of the AC field is $eV_{\mathrm{AC}}/\Gamma = 1$
  (broken line), $3$ (dotted line), and $5$ (solid line).
  }
  \label{fig:output_Single QD2}
\end{figure}

\subsubsection{Case of $\Gamma \gg \hbar \omega$}

In the adiabatic case of $\Gamma \gg \hbar \omega$, electrons do not
feel the oscillation of the energy level during the transport
through the QD.
Then, the electric current flows through the energy level
$\varepsilon_{1}(t)$,
\begin{equation}
I_{L \rightarrow R}(t) =
\frac{e}{\hbar}
\frac{\Gamma_L \Gamma_R}{\Gamma_L+\Gamma_R}
\left[ f_L(\varepsilon_{1}(t))
-f_R(\varepsilon_{1}(t)) \right].
\label{eq:seq_current_1QDb}
\end{equation}
The differential conductance is given by
\begin{equation}
\frac{d I_{L \rightarrow R}}{d \mu_L}(t) =
\frac{e}{\hbar}
\frac{\Gamma_L \Gamma_R}{\Gamma_L+\Gamma_R}
\frac{1}{4k_{\mathrm B} T}
\cosh^{-2}
\frac{\varepsilon_1(t)-\mu_{L}}
{2k_{\mathrm B} T},
\label{eq:seq_cond_1QDb}
\end{equation}
which oscillates by the time-dependent energy
level in Eq.\ (\ref{eq:F_AC}).

If the measurement time is much longer than the period
of the AC field, $2\pi/\omega$, the time-averaged conductance
$\overline{d I_{L \rightarrow R}/d \mu_L(t)}$ is observed. We plot
the value as a function of $\varepsilon_0$ in
Fig.\ \ref{fig:output_Single QD2}(a). The width of a single peak
around $\varepsilon_0=\mu_L$ is determined by the sum of
$k_{\mathrm{B}}T$ and the amplitude of the AC field, $eV_{\mathrm{AC}}$.

\subsection{Resonant tunneling}
\label{section:Calculated results1b}

Now we perform the summation of the perturbative series to the infinite order
to obtain the exact formula for the electric current. Following the
scattering theory,\cite{Hewson1993,JJSakurai2021}
we obtain the scattering matrix in an infinite series of $H_T$
from Eqs.\ (\ref{eq:wavefn-DQD1}) and (\ref{eq:wavefn-DQD2})
by the method of successive substitution.

\begin{figure}
  \centering
  \includegraphics*[width=0.65\textwidth]{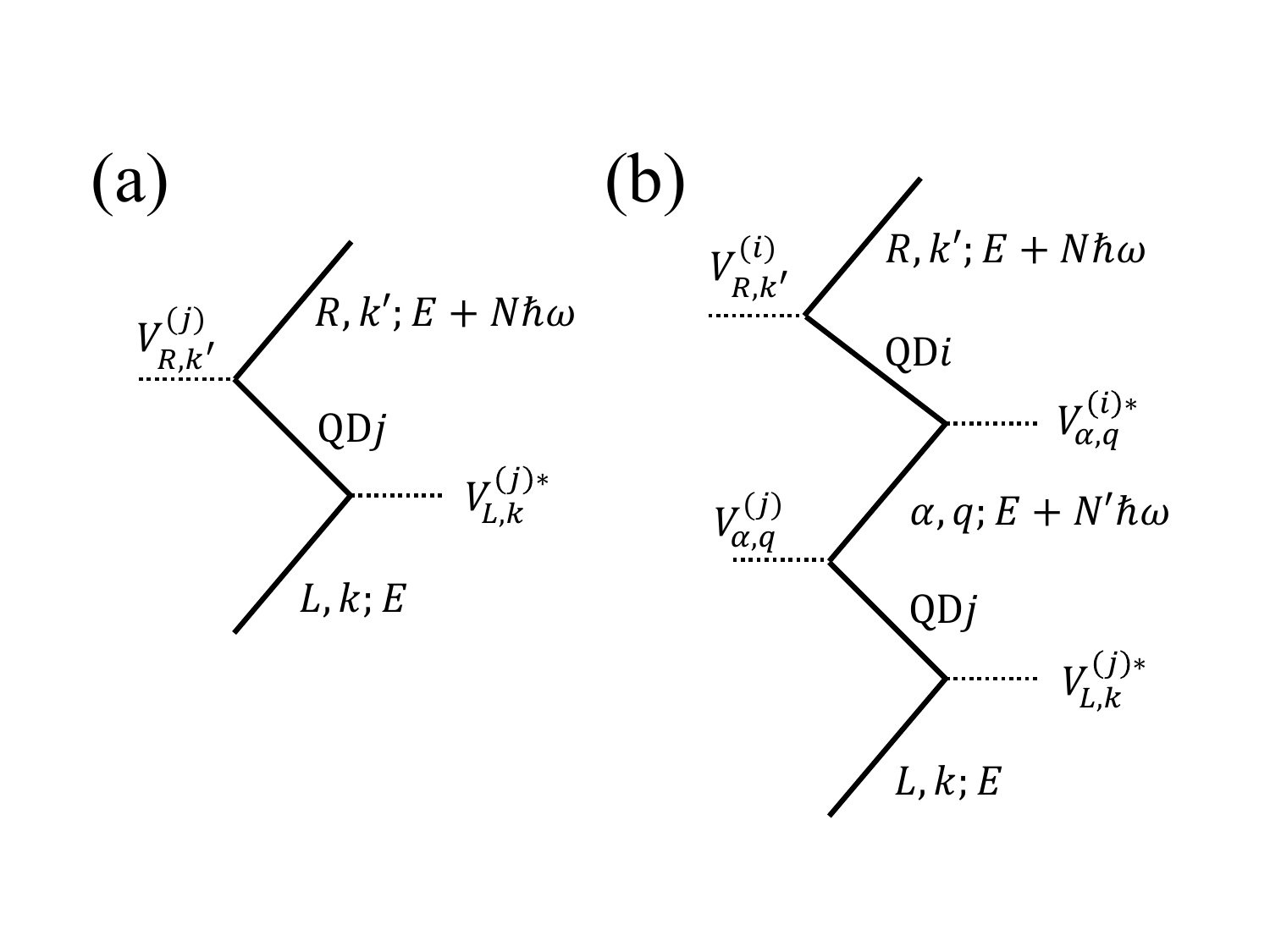}
  \caption{
  (a) First and (b) second terms of the perturbation series of the
  T-matrix with respect to the tunnel Hamiltonian $H_{\mathrm{T}}$:
  $V_{R,k'}^{(1)} G_{N,0}(E) V_{L,k}^{(1)*}$
  with $G_{N,0}(E)$ in Eq.\ (\ref{eq:GR-SingleQD}) for
  the single QD ($i=j=1$) and
  $\sum_{i,j=1,2}V_{R,k'}^{(i)} [G_{N,0}(E)]_{i,j}
  V_{L,k}^{(j) *}$ with $\mathbf{G}_{N,0}(E)$ in
  Eq.\ (\ref{eq:GR-DoubleQD}) for the DQD.
  $N$ is the number of photons in the final state.
  QD$j$ indicates the unperturbed Green's function
  of the QD, $G_{N_1,N_2}^{(0)}(E)$ in Eq.\ (\ref{eq:G0-SingleQD}),
  or that of the DQD, $\mathbf{G}_{N_1,N_2}^{(0)} (E)$ in Eq.\
  (\ref{eq:G0-DoubleQD}), where $N_1$ ($N_2$) is the number of
 photons  after (before) the propagation through the QD or DQD.
  $(\alpha,q; E)$ represents the
  propagator in lead $\alpha$, $1/(E-\varepsilon_q+i\eta)$.
 }
  \label{fig:Gdiagrams}
\end{figure}

\subsubsection{Case of $\Gamma \ll \hbar \omega$}

When $\Gamma \ll \hbar \omega$,
the scattering matrix from state $k$ in lead $L$ to
state $k'$ in lead $\alpha$ is given by
\begin{equation}
  S_{L,k \rightarrow \alpha,k'} =
  \delta_{\alpha, L} \delta_{k, k'}
  -2\pi i \sum_{N= -\infty}^{\infty}
  \langle \alpha, k' | T_N | L, k \rangle
  \delta(\varepsilon_{k'}-\varepsilon_k - N\hbar\omega).
  \label{eq:S-T-matrix}
\end{equation}
Here, $T_N$ is the T-matrix accompanied by the emission of $N$ photons
for $N>0$ or absorption of $|N|$ photons for $N<0$. It is written as
\begin{equation}
  \langle \alpha, k' | T_N | L, k \rangle
  = V_{\alpha,k'}^{(1)} G_{N,0} (\varepsilon_k) V_{L,k}^{(1)*},
\label{eq:T-matrix_1QD}
\end{equation}
with
\begin{equation}
    G_{N,0}(E) = G_{N,0}^{(0)}(E)
    -\frac{i}{2} \sum_{N'= -\infty}^{\infty}
   G_{N,N'}^{(0)}(E) \Gamma G_{N',0}^{(0)}(E) + \cdots
\label{eq:GR-SingleQD}
\end{equation}
and
\begin{equation}
G_{N,N'}^{(0)}(E) = \sum_{m}
\frac{J_{N-m}(\tilde{V}_{\mathrm{AC}}) J_{N'-m}(\tilde{V}_{\mathrm{AC}})}
{E - \varepsilon_{0}+ m\hbar\omega + i\eta}
\label{eq:G0-SingleQD}
\end{equation}
with a positive infinitesimal $\eta$.
$G_{N,N'}^{(0)}(E)=G_{N-N',0}^{(0)}(E+N'\hbar\omega)$ is the unperturbed
Green's function of the QD for an electron of energy $E$ with
$N'$ ($N$) photons before (after) the propagation through the QD.
The first and second terms of the infinite series in
the T-matrix, $V_{R,k'}^{(1)} G_{N,0}(E) V_{L,k}^{(1)*}$,
are schematically shown in Figs.\ \ref{fig:Gdiagrams}(a) and
\ref{fig:Gdiagrams}(b),
respectively. In the second term on the right side of Eq.\
(\ref{eq:GR-SingleQD}), the virtual state
$\Ket{\alpha,q}$ with $N'$ photons is taken into account by the
self-energy
$\Sigma_{\mathrm T}(E+N'\hbar\omega)=-i\Gamma/2$, where
\begin{equation}
\Sigma_{\mathrm T}(E)=\sum_{\alpha=L,R}\sum_{q}
V_{\alpha,q}^{(1)*}
\frac{1}{E-\varepsilon_q+i\eta}
V_{\alpha,q}^{(1)},
\end{equation}
assuming the wide-band limit in the leads.

Note that
the Green's function $G_{N,0}(E)$ is a component of the retarded
Green's function of the QD,
\begin{equation}
G^{\mathrm{r}}(t,t')=\frac{1}{i\hbar}
\langle \{ d(t), d^{\dag}(t') \} \rangle \theta(t-t'),
\label{eq:RetardedGreenFunc}
\end{equation}
with the step function $\theta(x)=1$ ($x>0$) and $0$ ($x<0$).
$G_{N,0}(E)$ is also related to the Green's function in the Floquet
theory, as shown in Appendix \ref{section:Gfn_Floquet}.

From Eq.\ (\ref{eq:GR-SingleQD}), we exactly obtain
\begin{equation}
G_{N,0} (E) = 
\sum_{m}
\frac{J_{N-m}(\tilde{V}_{\mathrm{AC}})J_{-m}(\tilde{V}_{\mathrm{AC}})}
{E - \varepsilon_{0}+ m\hbar\omega + i\Gamma/2}
\label{eq:GR-SingleQDa}
\end{equation}
using the relation of
$\sum_{N'} J_{N'-m}(\tilde{V}_{\mathrm{AC}})  J_{N'-n}(\tilde{V}_{\mathrm{AC}})
=J_{n-m}(0)=\delta_{m,n}$.
Using the transition rate from $\Ket{L,k}$ to $\Ket{R,k'}$,
\begin{equation}
w_{L,k \rightarrow R,k'}
=\frac{2\pi}{\hbar} \sum_{N}
|\langle R, k' | T_N | L, k \rangle|^2
\delta(\varepsilon_{k'}-\varepsilon_k - N\hbar\omega),
\label{eq:transition_by_T}
\end{equation}
and a similar expression for
$w_{R,k' \rightarrow L,k}$,\cite{Hewson1993,JJSakurai2021}
the electric current is written as
\begin{align}
I_{L \rightarrow R}
&=
e \sum_{k,k'}
\left\{
w_{L,k \rightarrow R,k'}f_L(\varepsilon_{k})
\left[ 1-f_R(\varepsilon_{k'}) \right]
-
w_{R,k' \rightarrow L,k}f_R(\varepsilon_{k'})
\left[ 1-f_L(\varepsilon_{k}) \right]
\right\}
\label{eq:current_org}
\\
&=\frac{e}{h} \Gamma_L \Gamma_R \sum_N \int d\varepsilon
\left| G_{N,0}(\varepsilon) \right|^2
\left[ f_L(\varepsilon)-f_R(\varepsilon) \right]
\nonumber
\\
&=\frac{e}{h} \Gamma_L \Gamma_R \sum_N \int d\varepsilon
\frac{\left[ J_{N}(\tilde{V}_{\mathrm{AC}}) \right]^2}
{(\varepsilon- \varepsilon_{0}+ N\hbar\omega)^2 + (\Gamma/2)^2}
\left[ f_L(\varepsilon)-f_R(\varepsilon)  \right].
\label{eq:current_1QD}
\end{align}
The same expression was obtained
previously.\cite{jauho_time-dependent_1994}
This yields the differential conductance
\begin{equation}
\frac{d I_{L \rightarrow R}}{d \mu_L}
=\frac{e}{h}
\frac{\Gamma_L \Gamma_R}{4k_{\mathrm B} T}
\sum_N \int d\varepsilon
\frac{\left[ J_{N}(\tilde{V}_{\mathrm{AC}}) \right]^2}
{(\varepsilon- \varepsilon_{0}+ N\hbar\omega)^2 + (\Gamma/2)^2}
\cosh^{-2}
\frac{\varepsilon-\mu_{L}}
{2k_{\mathrm B} T},
\label{eq:conductance_1QD}
\end{equation}
which is valid for any temperature $T$.
When $k_{\mathrm B} T \ll \Gamma$, it can be approximately
written as
\begin{equation}
\frac{d I_{L \rightarrow R}}{d \mu_L}
=\frac{e}{h} \Gamma_L \Gamma_R \sum_N
\frac{\left[ J_{N}(\tilde{V}_{\mathrm{AC}}) \right]^2}
{(\mu_L- \varepsilon_{0}+ N\hbar\omega)^2 + (\Gamma/2)^2}.
\label{eq:conductance_1QD2}
\end{equation}
This means resonant tunneling through the energy levels of the
polariton, $\varepsilon_{0}- N\hbar\omega$.
When $k_{\mathrm B} T \gg \Gamma$, Eq.\ (\ref{eq:conductance_1QD}) results
in the same expression as Eq.\ (\ref{eq:seq_cond_1QD}).
Hence, the treatment of the sequential transport is justified for
$k_{\mathrm B} T \gg \Gamma$.

In Fig.\ \ref{fig:output_Single QD}(b), we plot 
$d I_{L \rightarrow R} / d \mu_L$ in Eq.\ (\ref{eq:conductance_1QD2})
for $k_{B}T \ll \Gamma$ as a function of $\varepsilon_0$. 
We clearly observe subpeaks at $\mu_{L} = \varepsilon_{0} \pm \hbar\omega$
in addition to the main peak at $\mu_{L} = \varepsilon_{0}$.
With an increase in the amplitude of the AC field,
the height of the main peak decreases and that of the subpeaks increases.

\subsubsection{Case of $\Gamma \gg \hbar \omega$}

In the case of $\Gamma \gg \hbar \omega$,
the resonant tunneling takes place through the energy level
$\varepsilon_{1}(t)$. The electric current is given by
\begin{equation}
I_{L \rightarrow R}(t)
=\frac{e}{h} \Gamma_L \Gamma_R \int d\varepsilon
\frac{1}{[\varepsilon- \varepsilon_{1}(t)]^2 + (\Gamma/2)^2}
\left[ f_L(\varepsilon)-f_R(\varepsilon) \right].
\label{eq:current_1QDb}
\end{equation}
The differential conductance is
\begin{equation}
\frac{d I_{L \rightarrow R}}{d \mu_L}(t)
=\frac{e}{h}
\frac{\Gamma_L \Gamma_R}{4k_{\mathrm B} T}
\int d\varepsilon
\frac{1}{[\varepsilon- \varepsilon_{1}(t)]^2 + (\Gamma/2)^2}
\cosh^{-2}
\frac{\varepsilon-\mu_{L}}
{2k_{\mathrm B} T},
\label{eq:conductance_1QDb}
\end{equation}
which yields approximately
\begin{equation}
\frac{d I_{L \rightarrow R}}{d \mu_L}(t)
=\frac{e}{h}
\frac{\Gamma_L \Gamma_R}
{[\mu_L- \varepsilon_{1}(t)]^2 + (\Gamma/2)^2}
\label{eq:conductance_1QDb2}
\end{equation}
for $k_{\mathrm B} T \ll \Gamma$
and Eq.\ (\ref{eq:seq_cond_1QDb})
for $k_{\mathrm B} T \gg \Gamma$.

Figure \ref{fig:output_Single QD2}(b) presents
the time-averaged conductance
$\overline{d I_{L \rightarrow R}/d \mu_L(t)}$
for $k_{\mathrm B} T \ll \Gamma$
as a function of $\varepsilon_0$. We observe two
peaks around $\mu_L=\varepsilon_0 \pm
eV_{\mathrm{AC}}$, which are extreme values of
$F_{\mathrm{AC}}(t)$ in Eq.\ (\ref{eq:F_AC}).

\begin{figure}
  \centering
  \includegraphics*[width=0.4\textwidth]{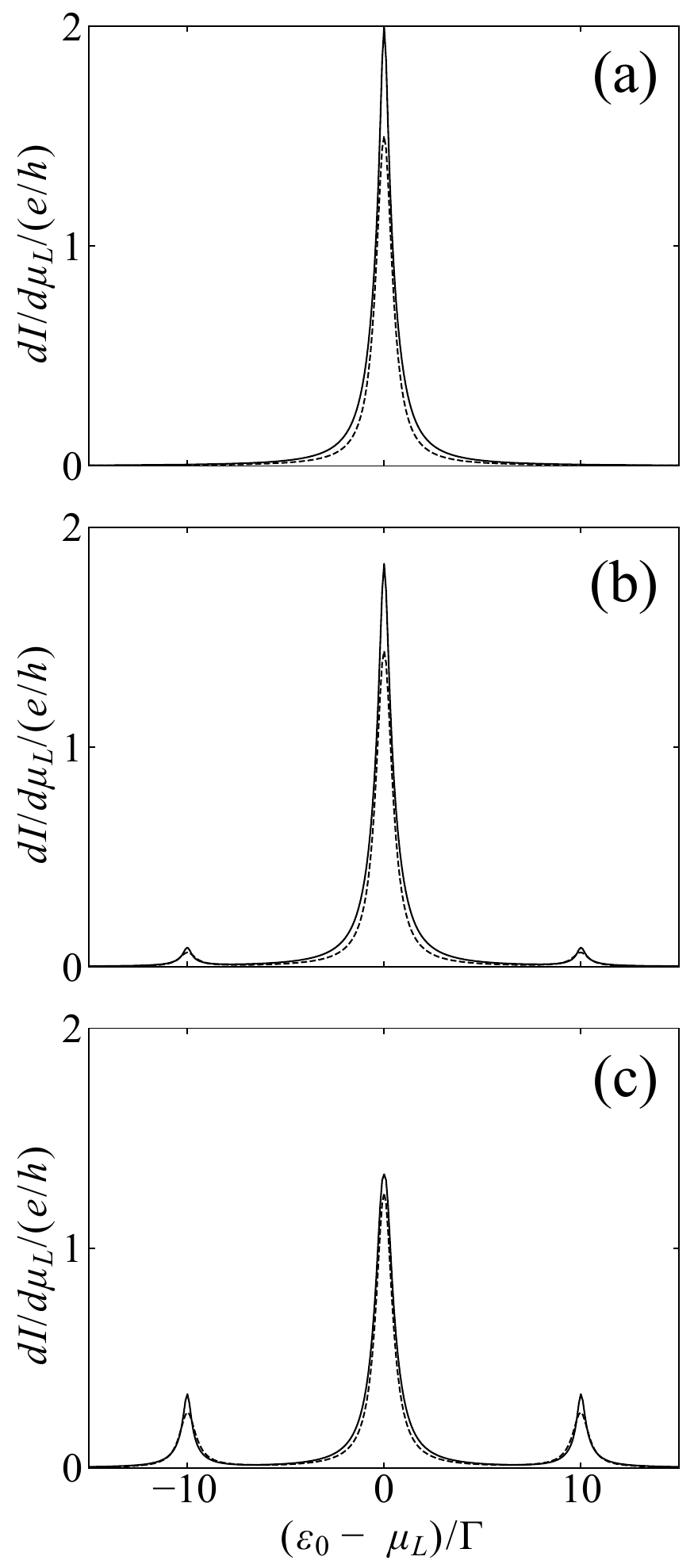}
  \caption{
  Differential conductance $d I_{L \rightarrow R}/d \mu_L$
  as a function of $\varepsilon_{0}$ for a DQD in the
  two-terminal geometry depicted in Fig.\ \ref{fig:model}(b).
  The energy level in QD1 is $\varepsilon_1(t)=F_{\mathrm{AC}}(t)$ in
  Eq.\ (\ref{eq:F_AC}) while that in QD2 is $\varepsilon_2=\varepsilon_0$.
  The linewidths due to the tunnel coupling to the leads are 
  $\Gamma_{jj}^{L} = \Gamma_{jj}^{R} \equiv \Gamma/2$ ($j=1,2$)
  and $p_{L} = p_{R} = 0.5$ in Eq.\ (\ref{eq:p}).
  The frequency of the AC field is $\hbar\omega/\Gamma = 10$
  and its amplitude is (a) $eV_{\mathrm{AC}}/\Gamma = 0$, (b) $5$, and
  (c) $10$.
  The AB phase for the magnetic flux penetrating between the QDs is
  $\phi=0$ (solid line) or $\pi$ (broken line). 
  }
  \label{fig:output_Double QD_shimpukuhenka}
\end{figure}

\section{Double Quantum Dot}
\label{section:Calculated results2}

In this section, we discuss the DQD models in Figs.\ \ref{fig:model}(b) and
\ref{fig:model}(c).
We restrict ourselves to the case of PAT for $\Gamma \ll \hbar \omega$.
As indicated in Appendix \ref{section:Calculation_suppl},
we obtain the scattering matrix
$S_{L,k \rightarrow \alpha,k'}$ 
as an infinite series of the perturbation in $H_T$, which is
related to the T-matrix by Eq.\ (\ref{eq:S-T-matrix}).

The T-matrix is common to the models in two- and three-terminal
geometries and is given by
\begin{equation}
  \langle \alpha, k' | T_N | L, k \rangle
  = \sum_{i,j=1,2}V_{\alpha,k'}^{(i)}
\left[ G_{N,0} (\varepsilon_k) \right]_{i,j}
V_{L,k}^{(j) *}
\label{eq:T-matrix_DQD}
\end{equation}
for the electron transfer accompanied by the emission
of $N$ photons ($N>0$) or the absorption of $|N|$ photons
($N<0$).
Here, $\left[ G_{N,0}(\varepsilon_k) \right]_{i,j}$ is
the $(i,j)$ component of 
$\mathbf{G}_{N,0}(\varepsilon_k)$ in a matrix form of $2\times2$
corresponding to QD1 and QD2,
\begin{align}
\mathbf{G}_{N,0}(E)
&= \mathbf{G}_{N,0}^{(0)}(E) -\frac{i}{2} \sum_{N'} \mathbf{G}_{N,N'}^{(0)}(E)
   \mathbf{\Gamma} \mathbf{G}_{N',0}^{(0)}(E) + \cdots 
\nonumber 
\\
&= \left[ \hat{\mathbf{G}^{(0)}}(E)
 \left[ \hat{\mathbf{1}} + \frac{i}{2} \hat{\mathbf{\Gamma}}
  \mathbf{G}^{(0)}(E) \right]^{-1} \right]_{N,0},
\label{eq:GR-DoubleQD}
\end{align}
using the unperturbed Green's function of the DQD,
\begin{equation}
\mathbf{G}_{N,N'}^{(0)} (E) =
\mathrm{diag} \left(
\sum_{m}
\frac{J_{N-m}(\tilde{V}_{\mathrm{AC}}) J_{N'-m}(\tilde{V}_{\mathrm{AC}})}
{E - \varepsilon_{0}+ m\hbar\omega + i\eta},
\frac{\delta_{N,N'}}{E - \varepsilon_0 + N\hbar\omega +i\eta}
\right).
\label{eq:G0-DoubleQD}
\end{equation}
In the second line of Eq.\ (\ref{eq:GR-DoubleQD}), we introduced
the supermatrices $\hat{\mathbf{G}^{(0)}}(E)$ and $\hat{\mathbf{\Gamma}}$,
the $(N,N')$ components of which are $\mathbf{G}_{N,N'}^{(0)} (E)$
and $\mathbf{\Gamma} \delta_{N,N'}$, respectively, for the compact form.

We cannot analytically sum up all the terms in the infinite series in Eq.\
(\ref{eq:GR-DoubleQD}) for the DQD as was done in the case of  a single QD
in Eq.\ (\ref{eq:GR-SingleQDa}).
Instead, we obtain the Green's function to the second order of
$\tilde{V}_{\mathrm{AC}}/2=eV_{\mathrm{AC}}/(2 \hbar \omega)$
in Appendix \ref{section:approxGreenF_DQD}.

In the two-terminal geometry in Fig.\ \ref{fig:model}(b),
Eq.\ (\ref{eq:current_org}) yields the electric current
\begin{align}
I_{L \rightarrow R}=
\frac{e}{h} \sum_{N} \int d\varepsilon
\Bigl\{
\mathrm{Tr} \left[ \mathbf{G}_{0,N}^{A}(\varepsilon) \mathbf{\Gamma}^{R}
\mathbf{G}_{N,0}(\varepsilon) \mathbf{\Gamma}^{L} \right]
f_{L}(\varepsilon) \left[ 1-f_{R}(\varepsilon + N\hbar\omega) \right]
\nonumber \\
+\mathrm{Tr} \left[ \mathbf{G}_{0,N}^{A}(\varepsilon) \mathbf{\Gamma}^{L}
\mathbf{G}_{N,0}(\varepsilon) \mathbf{\Gamma}^{R} \right]
f_{R}(\varepsilon) \left[ 1 - f_{L}(\varepsilon + N\hbar\omega) \right]
\Bigr\},
\label{eq:current-DoubleQD_exact}
\end{align}
where $\mathbf{G}_{0,N}^{A}(\varepsilon)$ is the adjoint matrix of
$\mathbf{G}_{N,0}(\varepsilon)$ in Eq.\ (\ref{eq:GR-DoubleQD}).

In the three-terminal geometry in Fig.\ \ref{fig:model}(c),
we set $\mu_{Ra}=\mu_{Rb} \equiv \mu_{R}$. In Eq.\ (\ref{eq:current_org}),
we take the summation over state $k'$ only in lead $Ra$ for the electric
current from lead $L$ to $Ra$, $I_{L \rightarrow Ra}$. Its expression
is given by Eq.\ (\ref{eq:current-DoubleQD_exact}) on replacing
$\mathbf{\Gamma}^{R}$ by $\mathbf{\Gamma}^{Ra}$.
We calculate $I_{L \rightarrow R}$ and $I_{L \rightarrow Ra}$ to the second
order in $\tilde{V}_{\mathrm{AC}}/2$.

\begin{figure}
  \centering
  \includegraphics*[width=0.43\textwidth]{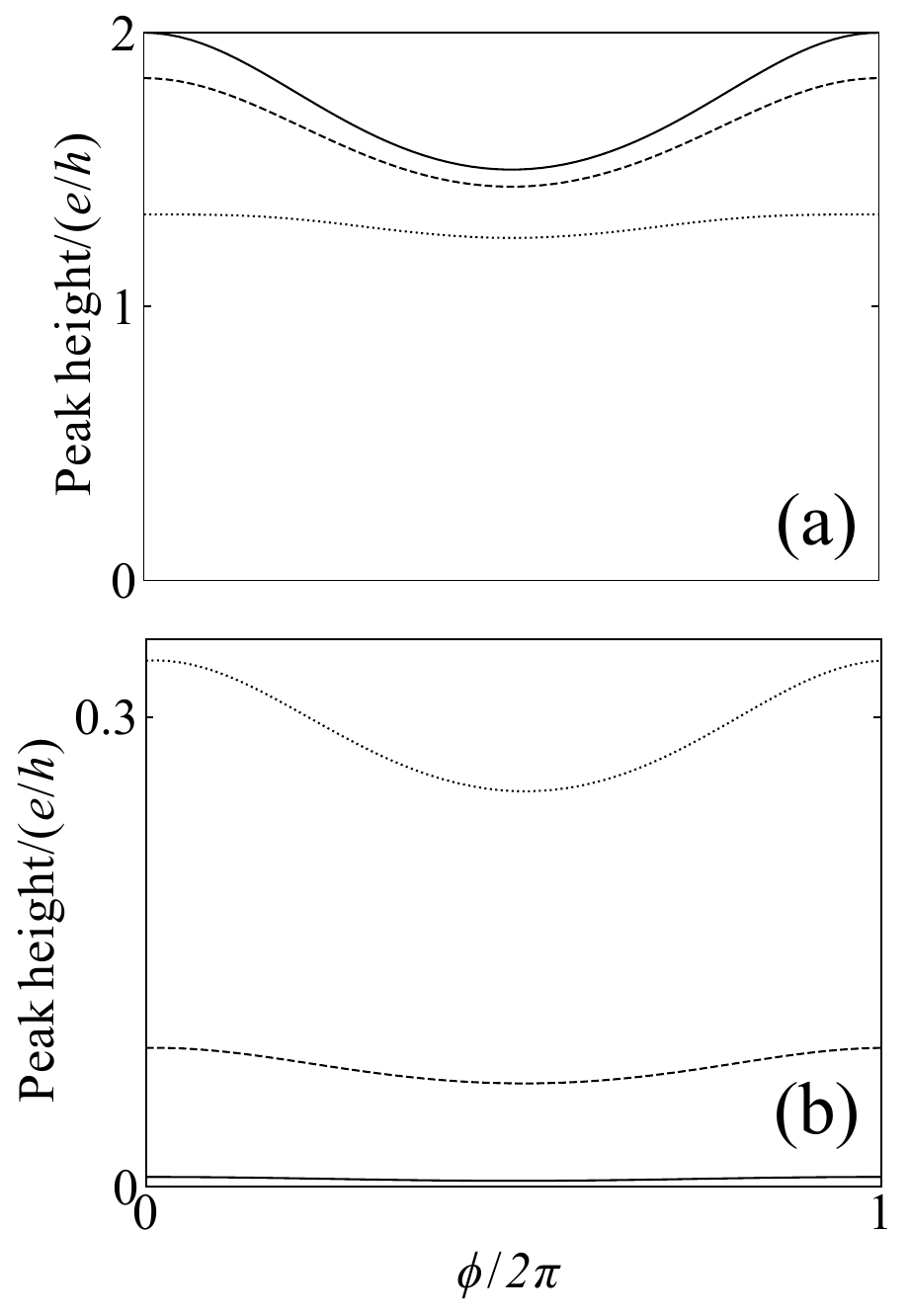}
  \caption{
  Differential conductance $d I_{L \rightarrow R}/d \mu_L$
  as a function of the AB phase $\phi$, i.e.,
  AB oscillation, for a DQD in the
  two-terminal geometry depicted in Fig.\ \ref{fig:model}(b),
  for (a) the main peak ($\varepsilon_{0}-\mu_{L} = 0$) and
  (b) the subpeaks ($\varepsilon_{0}-\mu_{L} = \pm \hbar\omega$).
  The energy levels in the QDs
  and the linewidth functions are the same as
  in Fig.\ \ref{fig:output_Double QD_shimpukuhenka}.
  The frequency of the AC field is $\hbar\omega/\Gamma = 10$
  and its amplitude is $eV_{\mathrm{AC}}/\Gamma = 0$
  (solid line), $5$ (broken line), and $10$ (dotted line). 
  }
  \label{fig:output_AB oscillation}
\end{figure}

\subsection{Double quantum dot with two terminals}

We discuss the calculated results for the PAT in the
DQD in the two-terminal geometry depicted in Fig.\ \ref{fig:model}(b).
The energy levels are
$\varepsilon_1(t)=F_{\mathrm{AC}}(t)$ in QD1 and
$\varepsilon_2=\varepsilon_0$ in QD2.
We set $\Gamma_{jj}^{L} = \Gamma_{jj}^{R} \equiv \Gamma/2$ ($j=1,2$)
and $p_{L} = p_{R} = 0.5$.

Figure \ref{fig:output_Double QD_shimpukuhenka} presents the
differential conductance $d I_{L \rightarrow R}/d \mu_L$
as a function of $\varepsilon_{0}$,
with the AB phase $\phi=0$ (solid line) or $\pi$ (broken line).
The amplitude of the AC field is (a) $eV_{\mathrm{AC}}/\Gamma = 0$,
(b) $5.0$, and (c) $10$.
In the absence of the AC field in
Fig.\ \ref{fig:output_Double QD_shimpukuhenka}(a),
the conductance peak is located at $\mu_L=\varepsilon_0$. The
AB effect causes the $\phi$ dependence of the peak height;
the positive (negative) interference effect increases (reduces)
the peak height for $\phi=0$ ($\pi$).
In the presence of the AC field in
Figs.\ \ref{fig:output_Double QD_shimpukuhenka}(b) and
\ref{fig:output_Double QD_shimpukuhenka}(c),
we observe the main peak at $\mu_L=\varepsilon_0$ and two
subpeaks at $\mu_L=\varepsilon_0 \pm \hbar \omega$.
The former decreases and the latter increases in height with
increasing AC field.
Note that both the main and subpeaks depend on the AB phase $\phi$.

We plot the AB oscillation of the main peak in
Fig.\ \ref{fig:output_AB oscillation}(a) and that of the subpeaks in
Fig.\ \ref{fig:output_AB oscillation}(b). The AC field suppresses
the amplitude of the oscillation at the main peak and enhances it at
the subpeaks.

The interference effect at the subpeaks can be understood as follows.
Consider the intermediate processes in the T-matrix
depicted in Fig.\ \ref{fig:Gdiagrams}.
When $N=\pm 1$, the electron transport is accompanied by the emission
or absorption of a photon. Then, the process in
Fig.\ \ref{fig:Gdiagrams}(a) does not involve the interference effect
between the QDs since the intermediate state is QD1 only.
In the process in Fig.\ \ref{fig:Gdiagrams}(b), on the other hand,
the interference exists when $N'=0$ or $N$ because an electron passes
by both the QDs in the first or second intermediate state, respectively.
Therefore, the interference effect is partly left for the subpeaks.
In the case of $N=0$ for the main peak, the interference exists
in the processes in Figs.\ \ref{fig:Gdiagrams}(a) and \ref{fig:Gdiagrams}(b)
with $N'=0$ but does not in that
in Fig.\ \ref{fig:Gdiagrams}(b) with $N' \ne 0$. In consequence,
the AB oscillation of the main peak is suppressed by the AC field.

\begin{figure}
\centering
\includegraphics*[width=0.43\textwidth]{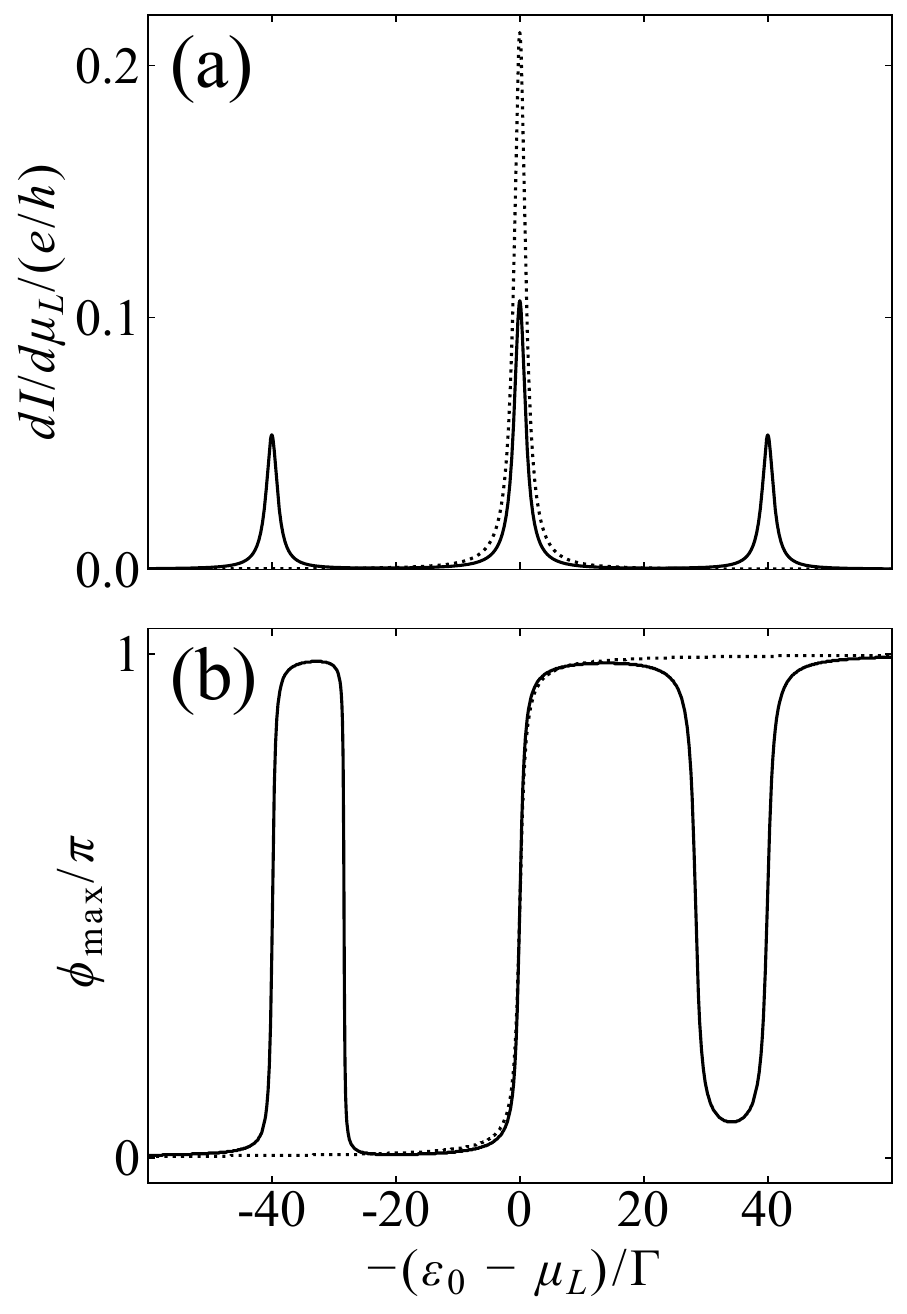}
\caption{
(a) Differential conductance $d I_{L \rightarrow Ra}/d \mu_L$
  as a function of $\varepsilon_{0}$ for a DQD in the
  three-terminal geometry depicted in Fig.\ \ref{fig:model}(c).
  The energy level in QD1 is $\varepsilon_1(t)=F_{\mathrm{AC}}(t)$
  in Eq.\ (\ref{eq:F_AC}) while that in QD2 is
  $(\varepsilon_{2} - \mu_{L})/\Gamma=70$.
  $\Gamma_{jj}^{L} = \Gamma_{jj}^{Ra} \equiv \Gamma/2$ ($j=1,2$)
  and $p_{L} = p_{Ra} = 0.5$ in Eq.\ (\ref{eq:p}).
  The frequency of the AC field is $\hbar\omega/\Gamma = 40$,
  whereas its amplitude is $eV_{\mathrm{AC}}/\Gamma = 40$ (solid line)
  and $0$ (dotted line).
  The tunneling coupling between QD1 and lead $Rb$ is 
  $\Gamma^{Rb}_{11} /(\Gamma^{Ra}_{11} + \Gamma^{Rb}_{11}) = 0.7$.
  The AB phase for the magnetic flux penetrating between the QDs
  is $\phi=0$.
(b) Measured phase shift $\phi_{\rm max}$ through QD1
  as a function of $\varepsilon_{0}$.
  $\phi_{\rm max}$ is the AB phase at which
  $d I_{L \rightarrow Ra}/d \mu_L$ is maximal for a given
  $\varepsilon_0$.
  Note that the abscissa shows $-\varepsilon_0$ in both panels.
}
\label{fig:output_phaseshift_peak}
\end{figure}

\subsection{Double quantum dot with three terminals}

We examine the PAT in the DQD in the three-terminal geometry depicted
in Fig.\ \ref{fig:model}(c), to demonstrate the phase measurement
through QD1 with $\varepsilon_1(t)=F_{\mathrm{AC}}(t)$
in Eq.\ (\ref{eq:F_AC}). We set the energy level $\varepsilon_2$
in QD2 far from $\mu_L$ [$(\varepsilon_{2} - \mu_{L})/\Gamma =70$]
to avoid the effect of its discreteness
on the resonant tunneling through the polariton states in QD1.
$\Gamma_{jj}^{L} = \Gamma_{jj}^{Ra} \equiv \Gamma/2$ ($j=1,2$)
and $p_{L} = p_{Ra} = 0.5$.

\begin{figure}
  \centering
  \includegraphics*[width=0.4\textwidth]{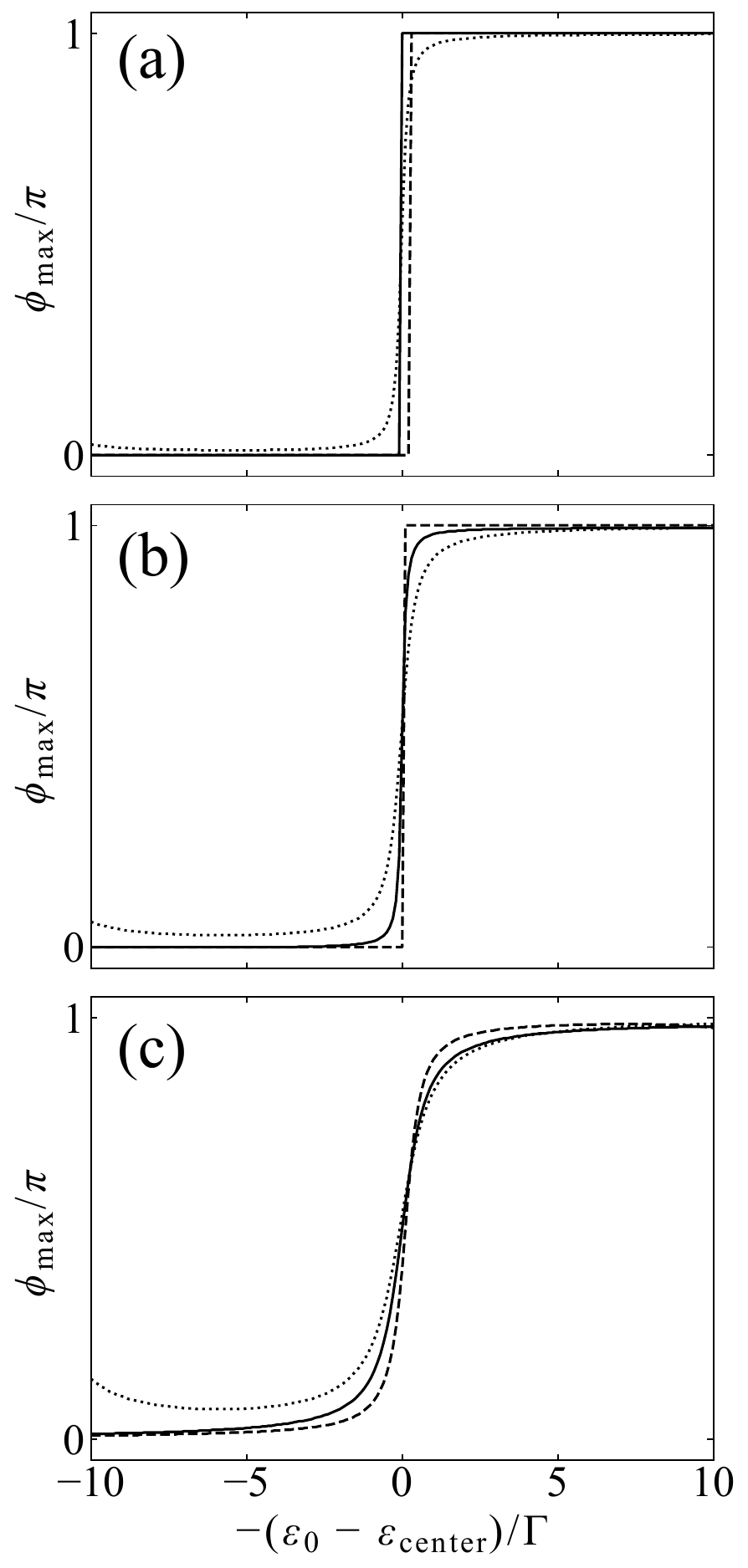}
  \caption{Measured phase shift $\phi_{\rm max}$ through
  QD1 in the DQD in the three-terminal geometry depicted in
  Fig.\ \ref{fig:model}(c), as a function of $\varepsilon_{0}$.
  The energy level in QD1 is
  $\varepsilon_1(t)=F_{\mathrm{AC}}(t)$ in
  Eq.\ (\ref{eq:F_AC}) while that in QD2 is
  $(\varepsilon_{2} - \mu_{L})/\Gamma =70$.
  $\phi_{\mathrm{max}}$ is plotted in the vicinity of the main peak at
  $\varepsilon_{\mathrm{center}}=\mu_L$ (solid line)
  and the subpeaks at
  $\varepsilon_{\mathrm{center}}=\mu_L-\hbar\omega$
  (broken line) and 
  $\varepsilon_{\mathrm{center}}=\mu_L+\hbar\omega$
  (dotted line).
  $\Gamma_{jj}^{L} = \Gamma_{jj}^{Ra} \equiv \Gamma/2$ ($j=1,2$)
  and $p_{L} = p_{Ra} = 0.5$ in Eq.\ (\ref{eq:p}).
  The frequency of the AC field is $\hbar\omega/\Gamma = 40$
  and its amplitude is $eV_{\mathrm{AC}}/\Gamma = 40$.
  The tunneling coupling between QD1 and lead $Rb$ is 
  (a) $\Gamma^{Rb}_{11} /(\Gamma^{Ra}_{11} + \Gamma^{Rb}_{11}) = 0$,
  (b) $0.4$, and (c) $0.7$.
  Note that the abscissa shows $-\varepsilon_0$ in all the panels.}
  \label{fig:output_phaseshift}
\end{figure}

Figure \ref{fig:output_phaseshift_peak}(a) shows the 
differential conductance $d I_{L \rightarrow Ra}/d \mu_L$
as a function of $\varepsilon_{0}$ with the AB phase $\phi=0$.
We choose a large value of $\hbar\omega$ to separate the
main and subpeaks ($\hbar \omega/\Gamma = 40$).
For comparison, we plot a single peak of
$d I_{L \rightarrow Ra}/d \mu_L$ in the absence of the AC
field with a dotted line.
The tunnel coupling to lead $Rb$ is fixed at
$\Gamma^{Rb}_{11} /(\Gamma^{Ra}_{11} + \Gamma^{Rb}_{11}) = 0.7$.
Note that the abscissa shows $-\varepsilon_0$ in
Figs.\ \ref{fig:output_phaseshift_peak} and
\ref{fig:output_phaseshift}.

Both the main peak and subpeaks change with the AB phase $\phi$,
indicating an AB oscillation.
We define the measured phase shift through QD1
as the AB phase $\phi_{\rm max}$
at which $d I_{L \rightarrow Ra}/d \mu_L$ is maximal for a given
$\varepsilon_0$. In Fig.\ \ref{fig:output_phaseshift_peak}(b),
we plot the measured phase as a function of $\varepsilon_0$.
We observe the change in $\phi_{\rm max}$ from 0 to $\pi$
around the subpeaks as well as the main peak with a decrease
in $\varepsilon_0$.
The change in $\phi_{\rm max}$ around the main peak is almost
the same as that around a single peak in the absence
of the AC field (dotted line).

Figure \ref{fig:output_phaseshift_peak}(b) also indicates
that $\phi_{\rm max}$ changes continuously
from $\pi$ to $0$ between a subpeak and the main peak, i.e.,
there is no ``phase lapse'' between the peaks.
Regarding the phase shift through a QD without the AC field,
the abrupt change between the successive conductance peaks
was reported\cite{schuster_phase_1997,
ji_phase_2000,Avinun-Kalish2005} and studied
theoretically.\cite{Karrasch2007}
This phase lapse takes place where the conductance
vanishes between the peaks.\cite{Karrasch2007}
In our case with the AC field, we observe neither the phase
lapse nor the zero point of $d I_{L \rightarrow Ra}/d \mu_L$
between the main peak and a subpeak.

We examine the role of the third lead, $Rb$, in the measurement
of the phase shift through QD1.
In Fig.\ \ref{fig:output_phaseshift}, we plot $\phi_{\rm max}$
in the vicinity of the main peak
($\varepsilon_{\mathrm{center}}=\mu_L$)
and subpeaks ($\varepsilon_{\mathrm{center}}=
\mu_L \pm \hbar \omega$), changing the tunnel coupling to
lead $Rb$ as (a) $\Gamma^{Rb}_{11}/(\Gamma^{Ra}_{11} +
\Gamma^{Rb}_{11}) = 0$, (b) $0.4$, and (c) $0.7$.
In the case of two terminals with $\Gamma^{Rb}_{11}=0$ in
Fig.\ \ref{fig:output_phaseshift}(a),
the measured phase $\phi_{\rm max}$ abruptly changes by $\pi$
at the peaks. ($\phi_{\rm max}$ seems to change gradually around one
of the subpeaks, which should be due to the effect from the
energy level $\varepsilon_{2}$ in QD2.\cite{comment1})
With an increase in the coupling to lead $Rb$, $\phi_{\rm max}$
changes more smoothly with $\varepsilon_0$ for all the peaks.

Finally, we discuss the shape of the measured phase $\phi_{\rm max}$
as a function of
$\varepsilon_{0}$ in Fig.\ \ref{fig:output_phaseshift}(c)
[and Fig.\ \ref{fig:output_phaseshift_peak}(b)] where the coupling
to the third lead is sufficiently large.
$\phi_{\rm max}$ behaves very similarly to one another around
the main peak and subpeaks. It is also almost identical to that
around a single peak in the absence of the AC field, as seen
in Fig.\ \ref{fig:output_phaseshift_peak}(b).
These functions of
$\phi_{\rm max}(\varepsilon_0 - \varepsilon_{\mathrm{center}})$
are qualitatively the same as $\theta_{\mathrm{QD}}$ defined by
\begin{equation}
\tan \theta_{\mathrm{QD}}=\frac{\Gamma_{1}/2}
{\varepsilon_0 - \varepsilon_{\mathrm{center}}},
\label{eq:theta_QD}
\end{equation}
where
$\Gamma_{1}=\Gamma^{L}_{11} + \Gamma^{Ra}_{11} + \Gamma^{Rb}_{11}$
is the total linewidth in QD1.
$\theta_{\mathrm{QD}}$ is the phase shift through QD1 by the
Breit--Wigner resonance when the coupling to QD2 is neglected.
The reason is given in Appendix \ref{section:phase_shift}.
If we replace $\Gamma_{1}$ and $\varepsilon_0$ in Eq.\
(\ref{eq:theta_QD}) by the renormalized values due to the
coupling to QD2, $\phi_{\rm max}$ would almost coincide with
$\theta_{\mathrm{QD}}$, as discussed previously in the absence
of the AC field.\cite{zhang_kondo_2022}

\section{Conclusions}
\label{section:Conclusion}

We calculated the electric current through a single QD and DQD
in an AC field by solving the time-dependent Schr\"odinger
equation using the scattering theory. The AC field is considered
as an oscillating energy level $F_{\mathrm{AC}}(t)$ in Eq.\
(\ref{eq:F_AC}) in a classical way.

For a single QD, we formulated the PAT when the linewidth
$\Gamma$ due to the tunnel coupling is much smaller than
$\hbar\omega$ and adiabatic transport when $\Gamma \gg \hbar\omega$.
In the case of PAT, we derived an analytical expression
for the electric current, which indicates the resonant tunneling
through the energy levels of polariton,
$\varepsilon_0+N\hbar\omega$ ($N=0,\pm 1, \pm 2, \cdots$).
In the case of adiabatic transport, the resonant tunneling
takes place through the energy level $\varepsilon(t)$.

For the DQD, we considered the situation that
one of the QDs (QD1) is irradiated by an AC field
while the other is not (QD2). We examined the PAT in the
presence of magnetic flux penetrating
between the QDs to elucidate the
coherent transport through the polariton states.
Using the infinite perturbation series with respect to
the tunnel Hamiltonian,
we calculated the electric current to the second order in
$eV_{\mathrm{AC}}/(2 \hbar \omega)$ in the two- and
three-terminal geometries.
We observed the AB effect both in the main peak
($N=0$) and in the subpeaks ($N \ne 0$). The coherence in the subpeaks
can be understood by the intermediate states in the T-matrix
depicted in Fig.\ \ref{fig:Gdiagrams}.
In the three-terminal geometry, we demonstrated the
measurement of the phase shift through QD1. The measured phase shift
changes continuously from 0 to $\pi$ around both
the main peak and subpeaks. No phase lapse is observed
between the peaks.

For the PAT through a single QD, the electric current in
Eq.\ (\ref{eq:seq_current_1QD})
was derived in Refs.\ \citen{platero_photon-assisted_2004} and
\citen{kohler_driven_2005} for $k_{\mathrm B}T \gg \Gamma$
and that in Eq.\ (\ref{eq:current_1QD}) was reported in Ref.\
\citen{jauho_time-dependent_1994}, using the Keldysh
nonequilibrium Green's function, Floquet theory, etc.
Compared with those methods, our calculation method based on
the scattering theory is helpful to understand the transport through a
time-dependent energy level intuitively on one hand, but
it is difficult to consider the electron--electron interaction
beyond the mean-field level on the other hand.

\section*{Acknowledgement}
This work was partially supported by JST SPRING,
Grant Number JPMJSP2123, and SUZUKI FOUNDATION.

\appendix

\section{Solution of time-dependent Schr\"odinger equation}
\label{section:Calculation_suppl}

For the DQD in Fig.\ \ref{fig:model}(b), the Hamiltonian is
given by Eq.\ (\ref{eq:schrodingereq}).
The substitution of the wavefunction in Eq.\ (\ref{eq:wavefn-DQD}) into
Eq.\ (\ref{eq:schrodingereq}) yields
\begin{align}
i \hbar \frac{d C_j}{dt} e^{-i E_j(t)/\hbar}
&=
\sum_{\alpha=L,R} \sum_{k} C_{\alpha,k}(t)
e^{-i \varepsilon_{k}(t-t_0)/\hbar}
  \left[ V_{\alpha k}^{(j)} \right]^*,
  \label{eq:wavefn-DQD1org}
\\
i \hbar \frac{d C_{\alpha,k}}{dt} e^{-i \varepsilon_{k}(t-t_0)/\hbar}
&=
\sum_{j=1,2} C_{j}(t) e^{-i E_j(t)/\hbar} V_{\alpha k}^{(j)}.
  \label{eq:wavefn-DQD2org}
\end{align}
Their integration from $t=t_0$ to $t_0+{\mathcal T}$ yields
\begin{align}
C_j(t_0+{\mathcal T})
&=
C_j(t_0)+
\sum_{\alpha=L,R} \sum_{k}
\frac{\left[ V_{\alpha k}^{(j)} \right]^*}{i\hbar} \int_{t_0}^{t_0+{\mathcal T}}
C_{\alpha,k}(t)
e^{-i[\varepsilon_k (t-t_0)-E_j(t)]/\hbar} dt,
\label{eq:wavefn-DQD1}
\\
C_{\alpha,k}(t_0+{\mathcal T})
&=
C_{\alpha,k}(t_0)+
\sum_{j=1,2} \frac{V_{\alpha k}^{(j)}}{i\hbar} \int_{t_0}^{t_0+{\mathcal T}}
C_j(t)
e^{i[\varepsilon_k (t-t_0)-E_j(t)]/\hbar} dt.
\label{eq:wavefn-DQD2}
\end{align}
From Eqs.\ (\ref{eq:wavefn-DQD1}) and (\ref{eq:wavefn-DQD2}), we derive the
transition rate to the lowest order in $H_{\mathrm{T}}$
in Sect.\ \ref{section:Calculation_suppl_1} and
the scattering matrix to the infinite order in $H_{\mathrm{T}}$
in Sect.\ \ref{section:Calculation_suppl_2}.

\subsection{Sequential tunneling through single QD}
\label{section:Calculation_suppl_1}

We examine the single QD model depicted in
Fig.\ \ref{fig:model}(a), by setting $V_{L k}^{(2)}=
V_{R k}^{(2)}=0$ and $C_{2}(t)=0$ in Eqs.\ (\ref{eq:wavefn-DQD1})
and (\ref{eq:wavefn-DQD2}).
We begin with the sequential tunneling to the lowest order in
$H_{\mathrm{T}}$.

We consider the transition of an electron from $\Ket{d_1}$
in the QD at $t=t_0$ to $\Ket{\alpha,k}$ in lead $\alpha$ at
$t=t_0+{\mathcal T}$.
For $C_1(t_0)=1$ and $C_{\alpha,k}(t_0)=0$, Eq.\ (\ref{eq:wavefn-DQD2})
gives
\begin{align}
C_{\alpha,k}(t_0+{\mathcal T})
& \simeq
\frac{V_{\alpha k}^{(1)}}{i\hbar} \int_{t_0}^{t_0+{\mathcal T}}
e^{-i(\varepsilon_0-\varepsilon_k)(t-t_0)/\hbar}
e^{-i\tilde{V}_{\mathrm AC}(\sin\omega t-\sin\omega t_0)} dt
\label{eq:tmp1}
\\
&= V_{\alpha k}^{(1)} \sum_{N=-\infty}^{\infty}
J_N(\tilde{V}_{\mathrm AC})
{\mathcal F}(\varepsilon_0-\varepsilon_k+N\hbar\omega)
e^{-i(N\omega t_0-\tilde{V}_{\mathrm AC}\sin\omega t_0)}
\label{eq:tmp2}
\end{align}
to the first order in $H_{\mathrm{T}}$, where
$\tilde{V}_{\mathrm AC}=eV_{\mathrm AC}/(\hbar\omega)$,
$J_N$ is the $N$th Bessel function, and
\begin{equation}
{\mathcal F}(E)=\frac{e^{-iE{\mathcal T}/\hbar}-1}{E}.
\label{eq:sincfn}
\end{equation}
Note that
$|{\mathcal F}(E)|^2=[4\sin^2 E{\mathcal T}/(2\hbar)]/E^2$
shows a peak at $E=0$ with the width of
$2\pi\hbar/{\mathcal T}$ and the height of
$({\mathcal T}/\hbar)^2$.
It has an asymptotic form of
$|{\mathcal F}(E)|^2 \sim 2\pi {\mathcal T} \delta(E)/\hbar$
for large ${\mathcal T}$.

\subsubsection{Case of $\Gamma \ll \hbar \omega$}

We set ${\mathcal T}$ to be the characteristic time of the
transport through the QD, ${\mathcal T} \sim \hbar/\Gamma$
(or several times of $\hbar/\Gamma$). Note that the transition
from $\Ket{d_1}$ to $\Ket{\alpha,k}$ takes place directly by the
perturbation $H_T$ in Eq.\ (\ref{eq:HT-DoubleQD}) in our model.
This is in contrast to the conventional situation of the scattering
theory where ${\mathcal T} \rightarrow \infty$ for an asymptotic transition
to the final state by the nonlocal perturbation.\cite{JJSakurai2021}

In the case of $\Gamma \ll \hbar \omega$, 
\begin{equation}
|C_{\alpha,k}(t_0+{\mathcal T})|^2
\simeq
\left| V_{\alpha k}^{(1)} \right|^2
\sum_N \left[ J_N(\tilde{V}_{\mathrm AC}) \right]^2
|{\mathcal F}(\varepsilon_0-\varepsilon_k+N\hbar\omega)|^2,
\end{equation}
since the peaks of $|{\mathcal F}(\varepsilon_0-\varepsilon_k+N\hbar\omega)|^2$
are separated from each other, and thus, the cross-terms in
$|C_{\alpha,k}(t_0+{\mathcal T})|^2$ (interference among different $N$'s)
can be disregarded.
Using the above-mentioned asymptotic form, the transition rate is given by
\begin{align}
w_{\mathrm{QD}\rightarrow \alpha,k}
&=
\frac{d}{d{\mathcal T}}|C_{\alpha,k}(t_0+{\mathcal T})|^2
\nonumber \\
&=
\frac{2\pi}{\hbar} \left| V_{\alpha k}^{(1)} \right|^2
\sum_N \left[ J_N(\tilde{V}_{\mathrm AC}) \right]^2
\delta(\varepsilon_0-\varepsilon_k+N\hbar\omega),
\label{eq:transition_rate_1QD}
\end{align}
which is an extension of the Fermi's golden rule.
As a result, the transition rate from the QD to lead $\alpha$
is
\begin{align}
\gamma_{\mathrm{QD}\rightarrow \alpha}
&=
\sum_k w_{\mathrm{QD}\rightarrow \alpha,k}
\left[ 1-f_{\alpha}(\varepsilon_k) \right]
\\
&=
\frac{\Gamma_{\alpha}}{\hbar}
\sum_N \left[ J_N(\tilde{V}_{\mathrm AC}) \right]^2
\left[ 1-f_{\alpha}(\varepsilon_0+N\hbar\omega) \right],
\end{align}
where $f_{\alpha}(E)=[1+e^{(E-\mu_{\alpha})/(k_{\mathrm B}T)}]^{-1}$ is
the Fermi distribution function in lead $\alpha$ at temperature $T$.
Similarly, the transition rate from lead $\alpha$ to the QD is given by
\begin{align}
\gamma_{\alpha \rightarrow \mathrm{QD}}
&=
\sum_k w_{\alpha,k \rightarrow \mathrm{QD}} f_{\alpha}(\varepsilon_k)
\\
&=
\frac{\Gamma_{\alpha}}{\hbar}
\sum_N \left[ J_N(\tilde{V}_{\mathrm AC}) \right]^2
f_{\alpha}(\varepsilon_0+N\hbar\omega).
\end{align}

The master equation for the probabilities $P_1$ for the occupied QD 
and $P_0$ for the empty QD is given by
\begin{equation}
\frac{d}{dt} P_1= -P_1 \sum_{\alpha=L,R}
\gamma_{\mathrm{QD}\rightarrow \alpha}
+P_0 \sum_{\alpha=L,R}
\gamma_{\alpha \rightarrow \mathrm{QD}},
\end{equation}
and $P_1+P_0=1$. In the stationary state ($dP_1/dt=0$),
the electric current is
\begin{align}
I_{L \rightarrow R}
&=
-e[P_1 \gamma_{\mathrm{QD}\rightarrow L}
 -P_0 \gamma_{L \rightarrow \mathrm{QD}}]
\nonumber \\
&=
e[P_1 \gamma_{\mathrm{QD}\rightarrow R}
 -P_0 \gamma_{R \rightarrow \mathrm{QD}}],
\end{align}
which yields Eq.\ (\ref{eq:seq_current_1QD}).

\subsubsection{Case of $\Gamma \gg \hbar \omega$}

In the adiabatic case of $\Gamma \gg \hbar \omega$, electrons do not
feel the oscillation of the energy level during
${\mathcal T} \sim \hbar/\Gamma$.
Then, $E_1(t) \simeq \varepsilon_{1}(t_0)(t-t_0)$
in Eq.\ (\ref{eq:E_1org}) for $t_0<t<t_0+{\mathcal T}$,
and thus, the electric current flows through the energy level
$\varepsilon_{1}(t_0)$.
The same procedure as in the previous subsection yields
\begin{equation}
I_{L \rightarrow R}(t_0) =
\frac{e}{\hbar}
\frac{\Gamma_L \Gamma_R}{\Gamma_L+\Gamma_R}
\left[ f_L(\varepsilon_{1}(t_0))
-f_R(\varepsilon_{1}(t_0)) \right].
\end{equation}
We replace $t_0$ by $t$ in Eq.\ (\ref{eq:seq_current_1QDb}).

\subsection{Resonant tunneling through single QD}
\label{section:Calculation_suppl_2}

Here, we perform the summation of the perturbative series to
the infinite order to obtain the exact formula for
the electric current in the case of a single QD.
Following the scattering
theory,\cite{Hewson1993,JJSakurai2021}
we consider an electron of state $k$ in lead $L$
at time $t=t_0-{\mathcal T}$ as the initial state, i.e.,
$C_1(t_0-{\mathcal T})=0$ and $C_{\alpha,k'}(t_0-{\mathcal T})=
\delta_{\alpha, L} \delta_{k, k'}$.
We evaluate the scattering matrix
\begin{equation}
  S_{L,k \rightarrow \alpha,k'} = C_{\alpha,k} (t_0+{\mathcal T})
  \label{eq:S-matrix-def}
\end{equation}
as an infinite series in $H_T$, using
Eqs.\ (\ref{eq:wavefn-DQD1}) and (\ref{eq:wavefn-DQD2})
by the method of successive substitution.

When $\Gamma \ll \hbar \omega$, we omit $t_0$,
which is irrelevant to this case.
The asymptotic form of ${\mathcal F}(E)$
in Eq.\ (\ref{eq:sincfn}) for large ${\mathcal T}$ results in
$S_{L,k \rightarrow \alpha,k'}$ in
Eqs.\ (\ref{eq:S-T-matrix}), (\ref{eq:T-matrix_1QD}), and
(\ref{eq:GR-SingleQD}).
The infinite series in Eq.\ (\ref{eq:GR-SingleQD}) can be
summed up analytically, which results in $G_{N,0}(E)$ in
Eq.\ (\ref{eq:GR-SingleQDa}).

When $\Gamma \gg \hbar \omega$, 
$E_1(t) \simeq \varepsilon_{1}(t_0)(t-t_0)$
in Eq.\ (\ref{eq:E_1org}) for $t_0<t<t_0+{\mathcal T}$.
The resonant tunneling through the energy level 
$\varepsilon_{1}(t_0)$ is derived in the same way.

\subsection{Resonant tunneling through DQD}
\label{section:Calculation_suppl_3}

In the case of the DQD in Figs.\ \ref{fig:model}(b) and
\ref{fig:model}(c), we consider the PAT for
$\Gamma \ll \hbar \omega$ only, setting $t_0=0$.
On the assumption that
$C_1(-{\mathcal T})=C_2(-{\mathcal T})=0$ and
$C_{\alpha,k'}(-{\mathcal T})=
\delta_{\alpha, L} \delta_{k', k}$ in the initial state,
we express the scattering matrix,
$S_{L,k \rightarrow \alpha,k'} = C_{\alpha,k} ({\mathcal T})$,
by an infinite series of the perturbation in $H_T$,
using Eqs.\ (\ref{eq:wavefn-DQD1}) and (\ref{eq:wavefn-DQD2})
by the method of successive substitution.
The result is given by Eqs.\ (\ref{eq:S-T-matrix}),
(\ref{eq:T-matrix_DQD}), and (\ref{eq:GR-DoubleQD}).

\section{Retarded Green's function and Floquet theory}
\label{section:Gfn_Floquet}

Here, we discuss the case of a single QD for simplicity.
The extension to the case of the DQD is straightforward.

Consider the retarded Green's function,
$G^{\mathrm{r}}(t,t')$ in Eq.\ (\ref{eq:RetardedGreenFunc}),
for the time-dependent Hamiltonian in Eq.\ (\ref{eq:H})
with $\varepsilon_2=0$ and $V_{L k}^{(2)}=V_{R k}^{(2)}=0$.
If we regard it as a function of $t_a=(t+t')/2$ and
$t_r=t-t'$,
\begin{equation}
G^{\mathrm{r}}(t_a+T,t_r)=G^{\mathrm{r}}(t_a,t_r)
\label{eq:GreenF_period}
\end{equation}
for $T=2\pi/\omega$ because of the periodicity of the Hamiltonian,
$H(t+T)=H(t)$, in the Floquet theory.\cite{floquet_sur_1883,
eliasson_floquet_1992,ishizuka_local_2017}.
Its Fourier transformation is given by
\begin{equation}
G^{\mathrm{r}}(t,t')=
\sum_{N=-\infty}^{\infty} \frac{1}{2\pi\hbar} \int dE
\bar{G}_N(E) e^{-i(N \omega t_a + E t_r/\hbar)}.
\label{eq:GreenF_Floquet}
\end{equation}

We introduce $G_{N,0}(E)$ in Eq.\ (\ref{eq:GR-SingleQD}), which
is related to $G^{\mathrm{r}}(t,t')$ in Eq.\
(\ref{eq:RetardedGreenFunc}) by
\begin{equation}
G^{\mathrm{r}}(t,t')=
\sum_{N=-\infty}^{\infty} \frac{1}{2\pi\hbar} \int dE
G_{N,0}(E) e^{-i[N \omega t + E (t-t')/\hbar]}.
\label{eq:GR-SingleQD2}
\end{equation}
A comparison with Eq.\ (\ref{eq:GreenF_Floquet}) results in
\begin{equation}
G_{N,0}(E)=\bar{G}_N(E+N\hbar\omega/2).
\end{equation}

Similarly to the unperturbed Green's function $G_{N,N'}^{(0)}(E)$
in Eq.\ (\ref{eq:G0-SingleQD}),
we define $G_{N,N'}$ for the propagator of electrons of
energy $E$ with $N'$ ($N$) photons before (after) the propagation
through the QD. It satisfies
\begin{align}
G_{N,N'}(E)
&=G_{N-N',0}(E+N'\hbar\omega)
\nonumber \\
&=
\bar{G}_{N-N'}(E+(N+N')\hbar\omega/2).
\end{align}

\section{Green's function to second order in $\tilde{V}_{\mathrm{AC}}/2$}
\label{section:approxGreenF_DQD}

From Eq.\ (\ref{eq:GR-DoubleQD}), we derive the expression of the Green's
function of the DQD, $\mathbf{G}_{N,0}(E)$, to the
second order in $\tilde{V}_{\mathrm{AC}}/2$.

Using the formula of the Bessel function,
$J_{0}(x) = 1 - (x/2)^2 + {\mathcal O}(x/2)^4$,
$J_{\pm 1}(x) = \pm x/2 + {\mathcal O}(x/2)^3$, and
$J_{\pm N}(x) = {\mathcal O}(x/2)^N$ ($N \ge 2$),
the unperturbed Green's functions in Eq.\ (\ref{eq:G0-DoubleQD})
are evaluated as
\begin{align}
  \mathbf{G}_{N,N}^{(0)}(E) &=
  \mathbf{A}_{N,N}(E) + \mathbf{B}_{N,N}(E),
\label{eq:G0-app-NN}
\\
\mathbf{A}_{N,N}(E) &=
\mathrm{diag} \left( g_{N}(E), h_{N}(E) \right)
\\
\mathbf{B}_{N,N}(E) &=
\left( \frac{\tilde{V}_{\mathrm{AC}}}{2} \right)^{2}
\mathrm{diag} \left( g_{N-1}(E) + g_{N+1}(E) -2 g_{0}(E), 0 \right)
\end{align}
for $N'=N$,
\begin{equation}
\mathbf{G}_{N,N+1}^{(0)}(E) = \mathbf{G}_{N+1,N}^{(0)}(E)
= \frac{\tilde{V}_{\mathrm{AC}}}{2}
\mathrm{diag} \left( g_{N}(E) - g_{N+1}(E), 0 \right)
\label{eq:G0-app-NN+1}
\end{equation}
for $N'=N \pm 1$, and
$\mathbf{G}_{N,N'}^{(0)} (E) =0$ otherwise, to the second order in
$\tilde{V}_{\mathrm{AC}}/2$. Here,
\begin{align}
  g_{N} \left( E \right) &= \frac{1}{E - \varepsilon_0 + N\hbar\omega + i\eta},
\\
  h_{N} \left( E \right) &= \frac{1}{E - \varepsilon_{2} + N\hbar\omega + i\eta}.
\end{align}
Substituting Eqs. (\ref{eq:G0-app-NN}) and (\ref{eq:G0-app-NN+1}) into Eq.\
(\ref{eq:GR-DoubleQD}), we obtain
\begin{align}
  \mathbf{G}_{0,0}(E) &\simeq
\mathbf{A}_{0,0}(E) \left[ \mathbf{1} + \frac{i}{2} \mathbf{\Gamma} \mathbf{A}_{0,0} \right]^{-1}
\notag \\
  &\qquad + \left[ \mathbf{1} +\frac{i}{2} \mathbf{A}_{0,0}(E) \mathbf{\Gamma} \right] ^{-1}
\mathbf{B}_{0,0}(E)
\left[ \mathbf{1} +\frac{i}{2} \mathbf{\Gamma} \mathbf{A}_{0,0}(E) \right] ^{-1}
\notag \\
  &\qquad - \frac{i}{2} \sum_{N'=\pm 1} \left[ \mathbf{1} +
\frac{i}{2} \mathbf{A}_{0,0}(E) \mathbf{\Gamma} \right] ^{-1}
\mathbf{G}_{0,N'}^{(0)} \mathbf{\Gamma} \mathbf{G}_{N',0}^{R},
  \label{eq:GR-app-00}
\\
  \mathbf{G}_{\pm 1,0}(E) &\simeq
\left[ \mathbf{1} +\frac{i}{2} \mathbf{A}_{\pm 1, \pm 1}(E) \mathbf{\Gamma} \right] ^{-1}
\mathbf{G}_{\pm 1,0}^{(0)}(E)
\left[ \mathbf{1} +\frac{i}{2} \mathbf{\Gamma} \mathbf{A}_{0,0}(E) \right] ^{-1}.
\label{eq:GR-app-10}
\end{align}

\section{Phase shift through QD in the AC field}
\label{section:phase_shift}

Let us consider the phase shift through QD1 in Fig.\
\ref{fig:model}(c), neglecting the coupling to QD2.
In the absence of the AC field, the phase shift of an
electron in the transfer from $\Ket{L,k}$ to $\Ket{R,k'}$
is given by the phase of
\begin{equation}
\langle R, k' | T | L, k \rangle
  = V_{R,k'}^{(1)}
  \frac{1}{\varepsilon_k-\varepsilon_0+i\Gamma_1/2} V_{L,k}^{(1)*},
\end{equation}
where
$\Gamma_{1}=\Gamma^{L}_{11} + \Gamma^{Ra}_{11} + \Gamma^{Rb}_{11}$
is the total linewidth in QD1.
This phase is equal to
$\tan^{-1}\Gamma_1/[2(\varepsilon_0-\varepsilon_k)]$
since $V_{R,k'}^{(1)}$ and $V_{L,k}^{(1)*}$ are real and
positive in our gauge.

In the presence of the AC field, the T-matrix is given by
Eq.\ (\ref{eq:T-matrix_1QD}) for the electron transfer
accompanied by the emission or absorption of $|N|$ photons,
with the Green's function in Eq.\ (\ref{eq:GR-SingleQDa})
if $\Gamma$ is replaced by $\Gamma_1$.
To the first order in $\tilde{V}_{\mathrm{AC}}/2$,
\begin{equation}
G_{0,0}(\varepsilon_k) \simeq
\frac{1}{\varepsilon_k-\varepsilon_0+i\Gamma_1/2}
\label{eq:G0approx}
\end{equation}
and
\begin{equation}
G_{\pm 1,0}(\varepsilon_k) \simeq
\pm \frac{\tilde{V}_{\mathrm{AC}}}{2}
\left(
\frac{1}{\varepsilon_k-\varepsilon_0+i\Gamma_1/2}
-\frac{1}{\varepsilon_k-\varepsilon_0 \pm \hbar \omega +i\Gamma_1/2}
\right).
\label{eq:G1approx}
\end{equation}
To the peak of the differential conductance
$d I_{L \rightarrow Ra}/d \mu_L$ at
$\mu_L=\varepsilon_0 + N \hbar\omega$ ($N=0, \pm 1$),
$\langle R, k' | T_{-N} | L, k \rangle$
mainly contributes when
$\hbar\omega \gg \Gamma_1$ and
$\tilde{V}_{\mathrm AC}=
eV_{\mathrm AC}/(\hbar\omega)$ is small.
Hence, the phase shift around the peak
is dominantly determined
by the phase of
$G_{0,0}(\varepsilon_k)$ in Eq.\
(\ref{eq:G0approx}) for $N=0$
and by that of the second term in $G_{-N,0}(\varepsilon_k)$
in Eq.\ (\ref{eq:G1approx}) for $N=\pm 1$.
This yields the same phase shift as in the absence of the
AC field if the origin of $\varepsilon_{0}$ is shifted by
$N\hbar\omega$.

In the transport experiment, the measured phase shift should be
identical to $\theta_{\mathrm{QD}}$ in Eq.\ (\ref{eq:theta_QD})
if the coupling to QD2 is negligibly small. In general, if
$\Gamma_{1}$ and $\varepsilon_0$ in Eq.\
(\ref{eq:theta_QD}) were replaced by the renormalized values
due to the coupling to QD2, $\phi_{\rm max}$ would almost
coincide with $\theta_{\mathrm{QD}}$, as discussed
previously in the case without the AC field.\cite{zhang_kondo_2022}


\begin{thebibliography}{10}
\bibitem{comment1}
In Fig.\ \ref{fig:output_phaseshift}(a) with $\varepsilon_{2} - \mu_{L} =70\Gamma$, $\phi_{\rm max}$ changes abruptly (gradually) around a subpeak at $\varepsilon_{0} - \mu_{L} =-\hbar\omega$ ($\hbar\omega$). If we choose $\varepsilon_{2} - \mu_{L} =-70\Gamma$, $\phi_{\rm max}$ changes abruptly (gradually) around a subpeak at $\varepsilon_{0} - \mu_{L} =\hbar\omega$ ($-\hbar\omega$).
\end{thebibliography}

\end{document}